\documentclass[aps,prx,reprint,longbibliography,superscriptaddress,floatfix, nofootinbib]{revtex4-2}
\usepackage{amsmath,amssymb,booktabs,graphicx,xcolor}
\usepackage{siunitx}
\usepackage[colorlinks=true,allcolors=blue!45!black]{hyperref}
\usepackage{footnote}
\usepackage{trimclip}
\graphicspath{{figs/}}
\newcommand{\fig}[1]{\includegraphics[width=\linewidth]{#1.pdf}}

\usepackage{upgreek}

\newcommand{\nhouba}[1]{{\color{black}#1}}

\usepackage{xspace}
\newcommand{\SFG}{SlotFlow-GB\xspace}

\begin{document}
\title{\SFG: \textcolor{black}{A Neural Framework for Trans-Dimensional Inference \\ of Overlapping Galactic Binaries}}
\author{Niklas Houba}
\email{nhouba@phys.ethz.ch}
\affiliation{Institute for Particle Physics and Astrophysics, ETH Z\"urich, 8093 Z\"urich, Switzerland}
\author{Giovanni Giarda}
\affiliation{Institute for Particle Physics and Astrophysics, ETH Z\"urich, 8093 Z\"urich, Switzerland}
\author{Davide Piras}
\affiliation{Institute for Particle Physics and Astrophysics, ETH Z\"urich, 8093 Z\"urich, Switzerland}
\affiliation{ETHZ Space Center, ETH Z\"urich, 8006 Z\"urich, Switzerland}

\begin{abstract}
The Laser Interferometer Space Antenna (LISA) will observe the millihertz
gravitational-wave sky, including \textcolor{black}{millions} of long-lived Galactic
binaries (GBs), tens of thousands of which are expected to be
individually resolvable throughout the mission's lifetime\textcolor{black}{.}
\textcolor{black}{Extracting these sources requires jointly determining their number
and physical parameters from strongly overlapping signals, posing
a computationally demanding trans-dimensional inference problem.}
We introduce \SFG, \textcolor{black}{a neural framework that constructs
a learned approximation to the trans-dimensional posterior,
producing a probability distribution over source multiplicity
and parameter densities for individual binaries in a single
evaluation, without requiring their number to be specified in advance.}
\textcolor{black}{We demonstrate this framework using year-long LISA simulations
containing up to ten binaries within a $2\,\upmu\mathrm{Hz}$
frequency interval, focusing on a high-SNR regime in which source
overlap, rather than individual-source detectability, is the
primary challenge.}
\textcolor{black}{The inferred source count differs from the true count by at most
one in $94\%$ of cases, while the model identifies individual
binaries even when their signals overlap strongly in frequency.}
\textcolor{black}{A complete evaluation, including parameter sampling, takes just}
$41$~ms per band on one NVIDIA GH200 GPU, or $530$~ms on eight
laptop CPU threads.
\textcolor{black}{We further demonstrate that these distributions can be used to initialize
subsequent likelihood-based inference, connecting fast amortized
predictions with joint parameter estimation.}
\end{abstract}

\maketitle
\flushbottom
\setcounter{topnumber}{3}\setcounter{dbltopnumber}{3}\setcounter{totalnumber}{4}
\renewcommand{\topfraction}{0.95}\renewcommand{\dbltopfraction}{0.95}
\renewcommand{\bottomfraction}{0.5}\renewcommand{\textfraction}{0.05}
\renewcommand{\floatpagefraction}{0.75}\renewcommand{\dblfloatpagefraction}{0.75}
\clubpenalty=10000 \widowpenalty=10000 \displaywidowpenalty=10000
\clubpenalties 3 10000 10000 150
\widowpenalties 3 10000 10000 150
\brokenpenalty=10000


\section{Introduction}\label{sec:intro}

The millihertz gravitational-wave sky contains a large population of Galactic binaries (GBs), tens of thousands of which are expected to be individually resolvable by the Laser Interferometer Space Antenna (LISA) mission~\cite{lisa, Lamberts:2019nyk}. Their signals are present simultaneously in the data, and those from nearby frequencies can overlap strongly. Recovering GB signals in a crowded band requires solving two coupled problems: determining how many binaries are present, and inferring the physical parameters of each one. Because the number of unknown parameters changes with the number of superposed GB signals, the inference task is intrinsically trans-dimensional.

Early studies showed that overlapping GBs could be recovered using blocked annealed
sampling with a variable number of sources~\cite{crowder} and reversible-jump Markov
chain Monte Carlo~\cite{green1995reversible, stroeer}. These methods were subsequently
developed into trans-dimensional global fits that jointly
infer multiple source populations and instrumental noise.
For instance, variable GB counts have been treated with
reversible-jump sampling~\cite{gbmcmc, glass, erebor}, while product-space model
selection has also been suggested for this problem~\cite{carlin1995bayesian, modular}. Other approaches
construct the catalog iteratively through repeated search and
thresholding~\cite{Strub:2024kbe}. Such analyses have been demonstrated on LISA Data Challenge simulated datasets~\cite{ldc}, including GPU-accelerated implementations for large-scale inference~\cite{erebor}.

Neural posterior estimation~\cite{papamakarios2016fast, pmlr-v97-greenberg19a, Cranmer:2019eaq} offers a different strategy. Instead of analyzing sources by exploring the likelihood, a neural estimator is trained on simulations to map the data directly to probability distributions over candidate sources and their parameters. This approach has become effective for current ground-based gravitational-wave observations, where compact-binary mergers can generally be treated as isolated events~\cite{Gabbard:2019rde, Green:2020hst}. The DINGO package, for example, produces posterior samples for detected signals from ground-based detectors in seconds~\cite{dingo, Dax:2022pxd}.

Space-borne gravitational-wave detectors such as LISA present a different regime. The data is expected to be signal dominated~\cite{Lamberts:2019nyk, karnesis}, so parameter estimation cannot generally be separated from determining how many sources are present. Neural posterior estimators have been developed for individual sources in LISA~\cite{Spadaro:2026evb} and, more recently, for pairs of overlapping GBs~\cite{delmond}. These models retain a fixed-dimensional inference target and therefore take the number of sources as given.

Inference over a variable number of components has been investigated in several settings within machine learning. Variational methods can represent variable-dimensional distributions~\cite{cosmic}; trans-dimensional flow-matching posterior estimation can target astronomical time
series~\cite{tfmpe}; and our earlier SlotFlow work combined slot-based representations with conditional normalizing flows for sinusoidal and Gaussian mixtures~\cite{slotflow}.

However, extending these ideas to GBs in LISA is substantially demanding: each source has eight physical parameters, and distinguishing signals that overlap strongly in frequency requires exploiting LISA's time- and sky-dependent response~\cite{cutler1998, Cornish:2002rt}.

\textcolor{black}{We introduce \SFG, a neural framework for end-to-end trans-dimensional inference that represents an unknown population of Galactic binaries through a distribution over source multiplicity and parameter densities for individual sources. It combines a fixed pool of competing source slots~\cite{slots}, with no predefined assignment to individual binaries; a permutation-invariant training loss that uses one-to-one matching to associate predicted slots with injected sources~\cite{detr}; and conditional normalizing flows~\cite{splines} that represent an eight-dimensional parameter density for each candidate. These densities retain correlations between individual source parameters, but not between different sources.}

We use this architecture to address four questions:
\begin{itemize}
    \item \textcolor{black}{Can one learned model estimate source counts and parameter distributions for an unknown number of overlapping GBs?}

    \item Which information in the LISA response allows binaries that strongly
    overlap in frequency to be distinguished?

    \item What is lost when the joint posterior is approximated by the marginal
    parameter densities of each source?

    \item Can the candidate sources identified by \SFG provide
    useful initial conditions for subsequent likelihood-based inference?
\end{itemize}

The paper is organized as follows. Section~\ref{sec:data} defines the inference problem and simulated data. Section~\ref{sec:methods} introduces the \SFG architecture and its probabilistic representation of the sources. Section~\ref{sec:results} presents several results, ranging from the evaluation of source recovery, source counting, and source parameter uncertainty to a demonstration of how \SFG can guide likelihood-based inference. Section~\ref{sec:remains} summarizes our conclusions and outlook.

\section{Problem setup}
\label{sec:data}

The goal of inference is to determine, from the superposition of all the GBs in a band, both how many are present and the physical parameters of each. For a band containing $K$ binaries, we write the observation as
\begin{equation}
d(t)
=
\sum_{i=1}^{K} h(t;\theta_i)+n(t),
\label{eq:model}
\end{equation}
where $d(t)$ denotes the data of a  given time-delay interferometry (TDI) channel, $h(t;\theta_i)$  is the LISA response to binary $i$, $n(t)$ is instrumental noise, and the number of sources $K$ is unknown.

Each binary is described by the eight-dimensional parameter vector
\begin{equation}
\theta_i=
\left(
f_{0,i},
\dot f_i,
\mathcal A_i,
\iota_i,
\lambda_i,
\beta_i,
\psi_i,
\phi_{0,i}
\right),
\end{equation}
containing its reference frequency $f_0$, frequency derivative $\dot f$, gravitational-wave amplitude $\mathcal A$, inclination $\iota$, ecliptic longitude $\lambda$ and latitude $\beta$, polarization angle $\psi$, and initial phase $\phi_0$.

The Bayesian target is
\begin{equation}
p\!\left(
K,\{\theta_i\}_{i=1}^{K}\mid d
\right),
\label{eq:target}
\end{equation}
whose dimension changes with $K$. Exchanging the labels of two binaries leaves the data unchanged, so
$\{\theta_1,\theta_2,\ldots,\theta_K\}$
and any permutation of the same elements represent the same configuration. A model should therefore recover the same sources regardless of how they are assigned to its output slots. The natural inference target is thus an \emph{unordered set} of sources whose number is unknown, which is the object \SFG is designed to approximate.

\subsection{Simulation domain}
\label{sec:simdomain}

We simulate $T_{\rm obs}=360$ days of LISA observations in the second-generation time-delay interferometry (TDI) channels $A$, $E$, and $T$~\cite{tinto_tdi,prince_ae}, adopting the equal-arm limit in which their instrumental noise is uncorrelated. Each observation contains $K$ GBs, with $K$ drawn uniformly from $\{0,\ldots,10\}$. Including observations without sources allows the network to learn when no binary is present. The binary reference frequencies $f_{0,i}$ are confined to a \SI{2}{\micro\hertz} region near \SI{3}{\milli\hertz}. We refer to one realization of the source population and instrumental noise as a \emph{scene}.

The source frequencies are drawn uniformly within \SI{\pm 1}{\micro\hertz} of a common center frequency. For each scene, this center is drawn uniformly within \SI{\pm 2}{\micro\hertz} of \SI{3}{\milli\hertz}. The network is not given the true center frequency. Instead, it receives an approximate center $f_c$, obtained by adding an independent uniform offset within \SI{\pm 30}{\nano\hertz} to the true center. This mimics the uncertainty of a preceding frequency-localization step, which we do not explicitly model. In contrast to a fixed band segmentation, however, the analyzed frequency window is not centered on $f_c$. Instead, $f_c$ guides a search for the strongest spectral peak, around which the window is then centered (Sec.~\ref{sec:repr}).

The remaining source parameters are drawn independently. The gravitational-wave amplitude $\mathcal A$ is sampled log-uniformly over $[10^{-22},10^{-20}]$, and the frequency derivative $\dot f$ log-uniformly over $[10^{-17},10^{-15}]\,\mathrm{Hz\,s^{-1}}$. Sky positions are isotropic, with ecliptic longitude $\lambda$ uniform on $[0,2\pi]$ and $\sin\beta$ uniform on $[-1,1]$. We draw $\cos\iota$ uniformly on $[-1,1]$, polarization angle $\psi$ uniformly on $[0,\pi]$, and initial phase $\phi_0$ uniformly on $[0,2\pi]$. These distributions define the source priors used for training and for the evaluation sets of
Sec.~\ref{sec:evalsets}, except where the latter deliberately fix individual parameters in order
to isolate a specific effect.

The amplitude range is chosen to produce a deliberately loud population rather than to reproduce the astrophysical GB population. This allows us to study source overlap and confusion in a regime where individual-source detectability is generally not the dominant limitation. 

We generate the GB waveforms with \texttt{jaxgb}~\cite{JAXGB:2026}, assuming circular binaries with constant frequency drift, and compute the LISA response using equal-armlength orbits from \texttt{lisaorbits}~\cite{bayle_2025_17296456}. The source frequencies occupy a \SI{2}{\micro\hertz} interval in each scene, but the waveforms are retained over a \SI{128}{\micro\hertz} band centered on \SI{3}{\milli\hertz}, which comfortably contains the full signal response. We add stationary Gaussian instrumental noise, generated independently in the Fourier domain for each TDI channel using the one-sided SciRDv1 power spectral densities implemented in \texttt{lisatools}~\cite{michael_katz_2026_18580841,LISA:SciRD2018}. The noise is simulated across the \SIrange{1}{5}{\milli\hertz} range, and we do not include an unresolved Galactic foreground. For the full $360$-day observation, the frequency-bin width is $\delta f_{\rm bin}=1/T_{\rm obs}=\SI{32.15}{\nano\hertz}$.

\begin{figure*}[!t]
\centering
\vspace{-5pt}
\fig{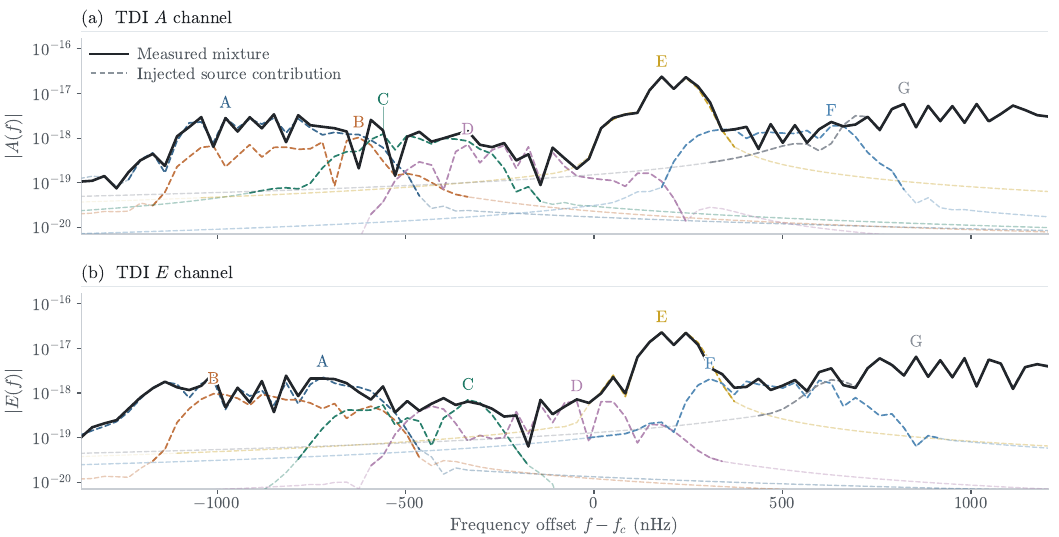}
\caption{Seven GBs overlapping within $\sim\SI{2}{\micro\hertz}$, shown in the TDI $A$ and $E$ channels. (a) Measured $A$-channel spectrum together with the seven injected contributions A--G, faded where each falls far below its own peak. (b) The same scene in the $E$ channel. The sky- and time-dependent LISA response weights each binary differently in the two channels. Source letters mark the peak of each binary in the respective channel. The same scene is followed in Figs.~\ref{fig:splash}, \ref{fig:showcase}, and \ref{fig:corner}.}
\label{fig:task}
\end{figure*}

We simulate $2\times10^6$ scenes and train on the first $98\%$ of them, holding out the
remainder as a validation split (Sec.~\ref{sec:evalsets}). Figure~\ref{fig:task} illustrates one such simulated scene containing seven overlapping GBs. Their different responses in the $A$ and $E$ channels provide information that helps distinguish sources with nearby frequencies.

\subsection{Evaluation sets}
\label{sec:evalsets}

We evaluate \SFG on three sets of simulated scenes, summarized in Table~\ref{tab:evalsets}. All use the waveform and noise model of Sec.~\ref{sec:simdomain} and differ in how the source parameters are sampled. The \emph{population set} measures source counting and parameter recovery across the training distribution. The \emph{two-source set} examines how frequency separation, angular separation, and amplitude contrast affect the recovery of overlapping binaries. The \emph{multiplicity grid} tests how recovery changes as the number of sources increases at approximately fixed frequency spacing.

The \emph{population set} is drawn from the held-out validation split of the training distribution. This split comprises the final $2\%$ of the $2\times10^6$ simulated scenes and is disjoint from the training data. We evaluate $20\,480$ scenes spanning the full range of source counts from zero to ten.

The \emph{two-source set} contains scenes with exactly two GBs. We characterize each pair by its frequency separation $\Delta f=|f_{0,1}-f_{0,2}|$, angular separation $\Delta\Omega=\arccos(\hat{\boldsymbol n}_1\cdot\hat{\boldsymbol n}_2)$, and amplitude ratio $R_{\mathcal A}=\mathcal A_{\rm faint}/\mathcal A_{\rm loud}\leq1$, where $\hat{\boldsymbol n}_i$ denotes the unit vector toward binary $i$, $A_{\rm faint}$ ($A_{\rm loud}$) is the amplitude of the fainter (louder) source, and smaller values of $R_{\mathcal A}$ correspond to larger differences in source amplitude.

We first consider a broadly sampled two-source set in which $\Delta f$ and $R_{\mathcal A}$ are drawn independently from log-uniform distributions and $\Delta\Omega$ from an isotropic angular distribution, rather than obtaining their joint distribution from the training prior. The frequency separation extends to \SI{4}{\micro\hertz}, beyond the maximum separation of \SI{2}{\micro\hertz} allowed by the training distribution. We then consider controlled configurations with prescribed values of $\Delta f$, $\Delta\Omega$, and $R_{\mathcal A}$: a grid spanning all three quantities and three sets of one-dimensional cuts in which one quantity is varied while the others are held fixed. These experiments allow us to examine the effects of frequency separation, sky separation, and relative source strength separately.

The \emph{multiplicity grid} test set increases the number of sources while keeping their characteristic frequency spacing approximately fixed. We consider $K\in\{2,3,5,10\}$, placing the source frequencies on a grid with nominal spacing \SI{200}{\nano\hertz} and a $\pm5\%$ jitter to avoid exactly periodic separations. The median minimum nearest-neighbor separation decreases only from \SI{200}{\nano\hertz} at $K=2$ to \SI{187}{\nano\hertz} at $K=10$. This construction allows us to study the effect of increasing source crowding without systematically reducing the characteristic separation between neighboring binaries.

We quote the optimal SNR of an individual binary, $\rho=\sqrt{\langle h,h\rangle}$, using the noise-weighted inner product defined in Appendix~\ref{app:assoc} with instrumental noise only. Other binaries in the scene and the unresolved Galactic foreground are not included as additional noise in this definition. In the broadly sampled two-source set, the median single-source SNR is $515$, with the central $90\%$ spanning approximately $49$--$5400$. The evaluated population also includes sources with SNRs as low as $\rho\approx30$.

\begin{table*}[t]
\caption{Overview of the evaluation sets. Here, $K$ is the number of binaries, $\Delta f$ and $\Delta\Omega$ are the frequency and angular separations, and $R_{\mathcal A}$ is the fainter-to-louder amplitude ratio. Numbers in parentheses indicate the number of sampled values of each parameter within the specified range. The table reports the number of scenes used in each evaluation.}
\label{tab:evalsets}
\begingroup
\small
\setlength{\tabcolsep}{2pt}
\begin{ruledtabular}
\begin{tabular}{llcrll}
Set & Sampling & $K$ & Scenes & Parameters & Used in \\ 
\colrule
Multiplicity & Spacing grid & $2,3,5,10$ &
\begin{tabular}[c]{@{}r@{}}$11\,500$\\ per $K$\end{tabular} &
\parbox[c]{6.cm}{Nominal spacing \SI{200}{\nano\hertz}, with $\pm5\%$ jitter} &
\parbox[c]{1.8cm}{Fig.~\ref{fig:recovery_mult}} \\
\colrule
Population & Validation & $0$--$10$ & $20\,480$ & \parbox[t]{6.7cm}{Training prior} & \parbox[t]{1.8cm}{Fig.~\ref{fig:conf}} \\
\colrule
Two-source & Broad sampling & $2$ & $24\,000$ & \parbox[t]{6.7cm}{\raggedright $\Delta f$: log-unif.\ \SI{1}{\nano\hertz}--\SI{4}{\micro\hertz}; $\Delta\Omega$: isotropic; $R_{\mathcal A}$: log-unif.\ $0.01$--$1$} & \parbox[t]{1.8cm}{Sec.~\ref{sec:calib}} \\
& Joint grid & $2$ & $22\,000$ & \parbox[t]{6.7cm}{\raggedright $\Delta f$ (6): \SI{3}{\nano\hertz}--\SI{1}{\micro\hertz}; $\Delta\Omega$ (5): $1^\circ$--$120^\circ$; $R_{\mathcal A}$ (3): $\{1,0.1,0.02\}$} & \parbox[t]{1.8cm}{Figs.~\ref{fig:res}, \ref{fig:tolerance}} \\
& $\Delta f$ cut & $2$ & $2700$ & \parbox[t]{6.7cm}{\raggedright $\Delta f$ (7): \SI{1}{\nano\hertz}--\SI{1}{\micro\hertz}; $\Delta\Omega=30^\circ$; $R_{\mathcal A}=1$} & \parbox[t]{1.8cm}{Fig.~\ref{fig:evidence}(a)} \\
& $\Delta\Omega$ cut & $2$ & $2300$ & \parbox[t]{6.7cm}{\raggedright $\Delta\Omega$ (6): $0.5^\circ$--$90^\circ$; $\Delta f=\SI{5}{\nano\hertz}$; $R_{\mathcal A}=1$} & \parbox[t]{1.8cm}{Fig.~\ref{fig:evidence}(b)} \\
& $R_{\mathcal A}$ cut & $2$ & $4300$ & \parbox[t]{6.7cm}{\raggedright $R_{\mathcal A}$ (6): $0.01$--$1$; $\Delta f\in\{32,320\}\,\mathrm{nHz}$; $\Delta\Omega=30^\circ$} & \parbox[t]{1.8cm}{Fig.~\ref{fig:evidence}(c)} \\ 
\end{tabular}
\end{ruledtabular}
\endgroup
\end{table*}

\section{Methods}
\label{sec:methods}

\begin{figure*}[t]
\centering
\fig{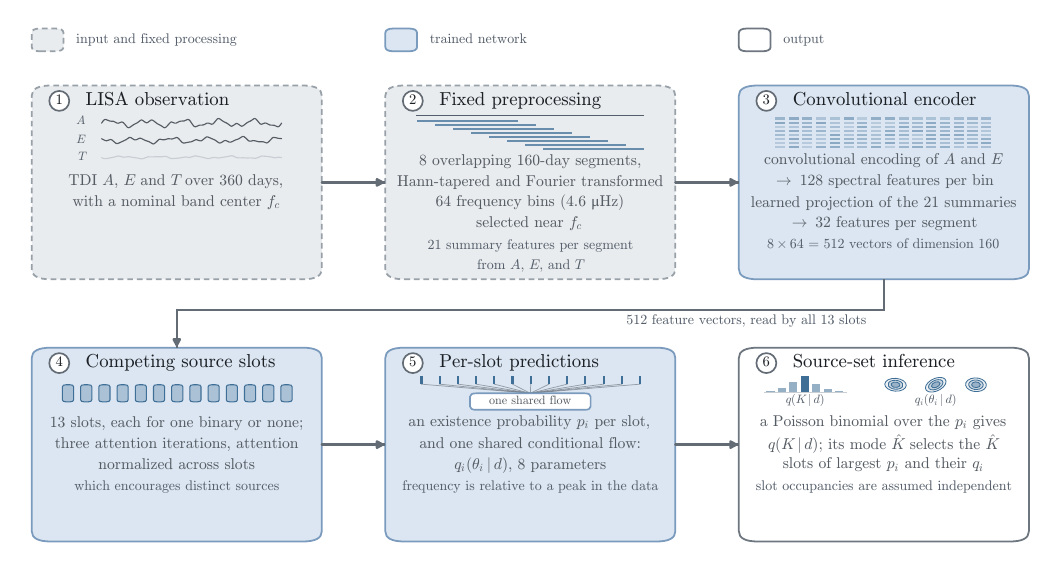}
\caption{Architecture of \SFG. The TDI observation and nominal frequency pointer (1) undergo fixed preprocessing, including time segmentation, spectral selection, and summary-feature extraction (2). A learned encoder combines convolutional features from the $A$ and $E$ channels with projected summary features from all three TDI channels (3). Competing source slots (4) predict existence probabilities and condition a shared normalizing flow for the source parameters (5). The resulting outputs comprise a source-count distribution $q(K\mid d)$ and the parameter densities $q_i(\theta_i\mid d)$ of the selected sources (6).
}
\label{fig:architecture}
\end{figure*}

\SFG maps a LISA observation directly to an unordered set of candidate binaries and their parameter densities in a single network evaluation, without performing a separate likelihood-based search. A fixed pool of candidate source slots represents the set, with each slot either describing one binary or remaining empty. Each slot predicts an existence probability and conditions a shared normalizing flow for the eight physical parameters of its candidate source. These predictions provide both a distribution over the number of binaries and parameter densities for the individual sources. Figure~\ref{fig:architecture} summarizes the six stages of the method.

\subsection{Data representation and competing source slots}
\label{sec:repr}
\label{sec:slots}

The input consists of the year-long TDI $A$, $E$, and $T$ observations and the nominal band center $f_c$ (Fig.~\ref{fig:architecture}, stage 1). As defined in Sec.~\ref{sec:simdomain}, $f_c$ identifies the approximate frequency region to be analyzed.

The fixed preprocessing stage (Fig.~\ref{fig:architecture}, stage 2) divides the $360$-day observation into eight overlapping $160$-day segments that together span the full observation. Each segment is Hann-tapered~\cite{blackman1958measurement} and Fourier transformed separately, giving a frequency-bin spacing of $1/T_{\rm seg}=72.3\,\mathrm{nHz}$, where $T_{\rm seg}=160$ days. This retains information about the time-dependent LISA response. To select the network input, we sum the $A$- and $E$-channel spectral amplitudes across segments, apply a Gaussian weighting centered on the nominal band center $f_c$, and select the frequency bin with the largest weighted amplitude. From each segment, we extract the same $64$ Fourier bins around this peak, corresponding to a bandwidth of $64/T_{\rm seg}\simeq\SI{4.6}{\micro\hertz}$. This bandwidth extends beyond the \SI{2}{\micro\hertz} interval of injected reference frequencies to reduce truncation of the binaries' time-dependent LISA responses near the boundaries of that interval, while keeping the network input compact. The same peak-search procedure is subsequently used with the slots' coarse frequency predictions to determine their source-specific frequency anchors (Sec.~\ref{sec:flow}). The fixed preprocessing also computes $21$ summary features per segment from the $A$, $E$, and $T$ channels, describing spectral power, inter-channel relationships, and phase evolution. 

A convolutional encoder (Fig.~\ref{fig:architecture}, stage 3) processes the selected $A$- and $E$-channel spectra using their amplitudes, phases, and relative channel responses, producing a $128$-dimensional feature vector at each frequency bin. The $21$ summary features from the fixed preprocessing are separately projected to $32$ dimensions and concatenated with the spectral features at each bin. This yields $8\times64=512$ feature vectors of dimension $160$, combining local spectral information with segment-level information from all three TDI channels. Learned segment and frequency-bin embeddings are added to encode each vector's position in the time--frequency representation. 
All architecture details are provided in Appendix~\ref{app:arch}.

The encoded observation is processed by $K_{\max}=13$ competing source slots (Fig.~\ref{fig:architecture}, stage 4), each represented by a $160$-dimensional vector. Through three iterations of slot attention~\cite{slots}, each slot attends to the encoding and updates its own representation. The attention weights are normalized across slots for each input feature, encouraging different slots to focus on different binaries rather than repeatedly representing the same source. Each slot starts from a distinct learned initialization, but no slot is assigned a fixed physical identity; the association between slots and binaries is determined separately for each observation.  Although the training scenes contain at most ten binaries, we use $13$ slots to provide additional candidates for source assignment. 

\subsection{Source parameter densities and source counting}
\label{sec:card}
\label{sec:flow}

Each slot predicts an existence probability $p_i$ and conditions a shared conditional spline flow~\cite{splines} (Fig.~\ref{fig:architecture}, stage 5). The flow models an eight-dimensional source parameter density $q_i(\theta_i\mid d)$, with the same flow weights used for all slots and each slot's learned representation providing its conditioning information. Auxiliary prediction heads estimate sky direction and frequency: the sky estimates are used to associate slots with injected binaries during training, while the frequency estimates guide the localization of individual sources in the observed spectrum. The flow represents the uncertainty and correlations among the parameters of each binary, with the physical frequency reconstructed as described below. The flow architecture and auxiliary prediction heads are detailed in Appendix~\ref{app:arch}.

We infer each binary's frequency relative to a reference frequency obtained from the observed spectrum. Each slot first predicts an approximate frequency, which guides the peak-search procedure introduced in Sec.~\ref{sec:repr}. The frequency of the selected spectral bin defines the slot-specific reference $f_i^{\rm a}$. 
A regression head then predicts a normalized frequency offset $\hat f_i$, while the flow models the remaining residual $\delta_i$, giving
\begin{equation}
f_{0,i}=f_i^{\rm a}+F(\hat f_i+\delta_i),
\qquad F=800\,\mathrm{nHz}.
\label{eq:F}
\end{equation}
Here, $F$ is a fixed scale used to normalize the frequency offsets during training. This parameterization allows the flow to model frequency deviations locally around a data-derived reference. Different binaries may select the same reference bin but still have distinct inferred frequencies.

The per-slot predictions are combined to obtain the source-count distribution and the inferred source set (Fig.~\ref{fig:architecture}, stage 6). We treat each slot as independently occupied or empty, with occupation probability $p_i$ determined by the observation. The resulting distribution over the total number of occupied slots is a Poisson binomial distribution, which allows different occupation probabilities for different slots. Its probability-generating polynomial is
\begin{equation}
G(x)=\prod_{i=1}^{K_{\max}}(1-p_i+p_i x),
\label{eq:pgf}
\end{equation}
where $x$ is an auxiliary variable used to count occupied slots. In each factor, $1-p_i$ represents an empty slot and $p_i x$ an occupied slot. Expanding the product combines all possible occupancy patterns, with each occupied slot contributing one power of $x$.\footnote{As an example, for two slots, $G(x)=(1-p_1)(1-p_2)+[p_1(1-p_2)+(1-p_1)p_2]x+p_1p_2x^2$. The coefficients of $x^0$, $x^1$, and $x^2$ give the probabilities of zero, one, and two sources, respectively.} The probability of exactly $K$ sources is therefore
\begin{equation}
q(K\mid d)=[x^K]G(x),
\label{eq:poissonbinomial}
\end{equation}
where $[x^K]G(x)$ denotes the coefficient of $x^K$ in $G(x)$. This calculation is exact for the predicted slot probabilities under the assumed independence of their occupancies and defines the model's approximation to the marginal count posterior in Eq.~\eqref{eq:target}, retaining uncertainty over the number of sources.

To obtain the inferred source set, we select the most probable count $\hat K$ under $q(K\mid d)$ and retain the $\hat K$ slots with the largest existence probabilities. Each selected slot is reported together with its eight-dimensional density $q_i(\theta_i\mid d)$, preserving the uncertainty and parameter correlations within that source. Stage 6 of Fig.~\ref{fig:architecture} illustrates both outputs: the full source-count distribution and the parameter densities of the selected sources. The inferred set of source parameters is used for the recovery metrics of Sec.~\ref{sec:works} and for initializing the likelihood-based analysis of Sec.~\ref{sec:globalfit}.

The existence probabilities and parameter densities together describe the network's uncertainty about which slots contain binaries and what their parameters are. For each slot $i$, we introduce a binary variable $b_i$, which equals one if the slot is occupied and zero if it is empty. An empty slot contributes a probability $1-p_i$, while an occupied slot contributes its existence probability $p_i$ multiplied by the parameter density $q_i(\theta_i\mid d)$. Combining these contributions across all slots gives
\begin{equation}
q(\mathbf b,\boldsymbol\theta_{\mathbf b}\mid d)
=\prod_{i=1}^{K_{\max}}(1-p_i)^{1-b_i}\left[p_iq_i(\theta_i\mid d)\right]^{b_i},
\label{eq:factorized}
\end{equation}
where $\mathbf b=(b_1,\ldots,b_{K_{\max}})$ specifies which slots are occupied and $\boldsymbol\theta_{\mathbf b}=\{\theta_i:b_i=1\}$ contains the parameters of those slots. The exponents ensure that each slot contributes only the factor corresponding to its occupancy. The flow captures correlations among the eight parameters of an individual binary, but the product assumes that different binaries' parameter densities are independent given the observation. For example, the inferred parameters of two overlapping binaries may be correlated because changes in one waveform can be compensated by changes in the other. Such cross-source correlations are not represented by the network's factorized density and must be explored through subsequent joint likelihood-based inference.

\begin{figure*}[t]
\centering
\fig{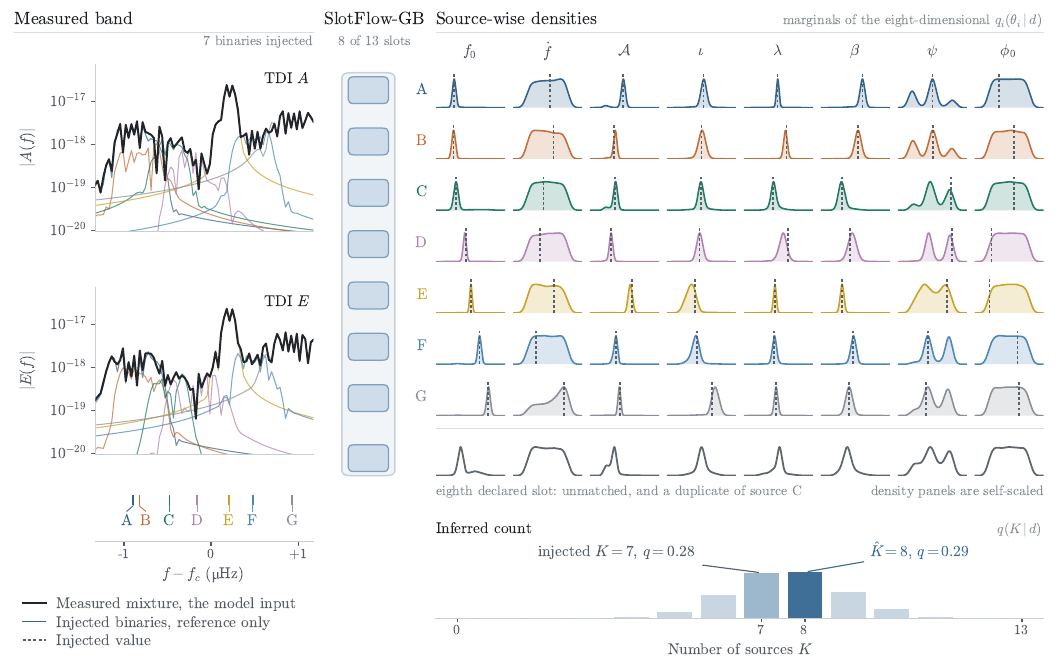}
\caption{\SFG applied to the seven-binary scene of Fig.~\ref{fig:task}. Seven binaries are injected, while the model reports $\hat K=8$ sources. All eight selected slots are shown; seven are labeled A--G according to their associated injections, and the unmatched slot is unfilled. Left: the measured TDI $A$ and $E$ mixtures, with individual injected signals shown dashed for reference only. Center: the selected source slots. Right: one-dimensional marginals of their eight-dimensional densities $q_i(\theta_i\mid d)$ and the inferred count distribution $q(K\mid d)$. The flow represents a joint eight-dimensional density for each source, not eight independent one-dimensional densities.}
\label{fig:splash}
\end{figure*}

Figure~\ref{fig:splash} illustrates the predictions for the example observation of Fig.~\ref{fig:task} containing seven overlapping GBs.

\subsection{Training}
\label{sec:loss}

We train \SFG on the first $98\%$ of the $2\times10^6$ simulated scenes described in Sec.~\ref{sec:simdomain}, reserving the remaining $2\%$ ($40\,000$ scenes) for validation. This held-out subset also serves as the population evaluation set in Sec.~\ref{sec:evalsets}. Each scene contains between zero and ten injected binaries. Training runs for $183$ epochs over the $1.96\times10^6$ training scenes, using eight GH200 GPUs with a batch size of $64$ per GPU, corresponding to approximately $7\times10^5$ optimizer steps. We use Adam \cite{Adam} with an initial learning rate of $4\times10^{-4}$. Validation is performed every two epochs, and the learning rate is halved if the exact-count accuracy does not improve over six consecutive validation checks. By the end of training, the learning rate has decreased by a factor of $128$. We stop training after the validation metrics plateau. Over the final $20$ epochs, the validation log density varies by less than $0.03$ and the exact-count accuracy by less than $0.01$, compared with a total improvement of $7.3$ in log density over the full training run.
 
Because the injected binaries have no intrinsic ordering, we associate them with the predicted slots using one-to-one Hungarian matching~\cite{detr,kuhn1955hungarian}. The matching cost is the angular separation between each slot's auxiliary sky-direction prediction (Sec.~\ref{sec:card}) and the injected sky direction. We choose this cost because the sky head can be supervised directly from the start of training, whereas the flow densities are initially similar across slots and provide little information for distinguishing individual sources. The matching assigns each injected binary to a different slot, with any remaining slots treated as empty. These assignments determine the training targets but do not impose fixed source identities across observations. Since the matching determines which source supervises each slot, the choice of cost may affect the network's ability to resolve overlapping sources. We have not compared against a cost based on the negative log density assigned by the flow to each injected source, which would use all eight source parameters rather than the sky direction alone. The resolution limits reported in Sec.~\ref{sec:disentangle} therefore apply to the sky-based matching used here. This approach already achieves the source recovery and counting performance reported in Sec.~\ref{sec:results}.

The network is trained to estimate the physical parameters of each binary, predict which slots contain sources, and learn the sky-position and frequency estimates used for source assignment and frequency reconstruction. For each matched slot, we minimize the negative log density assigned by its conditional flow to the injected binary's physical parameters, encouraging the predicted distribution to concentrate on the correct source parameters. The frequency is treated according to Eq.~\eqref{eq:F}: the flow models the distribution of the residual $\delta_i$, while a separate regression head predicts the normalized frequency offset $\hat f_i$ from the reference frequency $f_i^{\rm a}$. The initial frequency prediction, which guides the search for this reference in the observed spectrum, is also trained against the injected frequency. Similarly, the predicted sky direction is trained to agree with the injected position, providing the information needed for matching binaries to slots. Finally, the existence probability $p_i$ is trained toward one for slots assigned to injected binaries and toward zero for empty slots. Unmatched slots receive no source-parameter or auxiliary regression targets, since they do not correspond to an injected binary. 

The total loss function is composed of several pieces, which we describe in detail in Appendix~\ref{app:arch} together with the regularization terms, relative loss weights, and implementation details. The training losses encourage the network to associate injected binaries with distinct slots, distinguish occupied from empty slots, and estimate the physical parameters and uncertainties of the recovered sources.

\subsection{Source association and evaluation metrics}
\label{sec:works}
\label{sec:examples}

To evaluate how well \SFG recovers the injected binaries, we first determine which reported sources correspond to which injections. We consider two association criteria: one based on the difference between inferred and injected frequencies, and another based on the agreement between their corresponding gravitational-wave waveforms. 
Both methods ensure that each reported source is associated with at most one injected binary and vice versa, preventing multiple predictions from being counted as separate recoveries of the same source.

The first approach, \emph{frequency matching}, associates sources according to their inferred frequencies. For each reported source, we draw samples from its learned parameter density and select the sample with the highest predicted density as a representative parameter estimate. We compare its physical frequency $f_0$ with those of the injected binaries and accept an association when the difference is at most $500\,\mathrm{nHz}$, corresponding to approximately seven Fourier bins of the segmented observation. This tolerance provides a practical measure of source localization within the observed frequency band. For successfully associated binaries, we compare the parameters of the selected sample with the injected values to quantify the accuracy of the recovered source parameters.

The second approach, \emph{waveform association}, determines whether a reported source corresponds to an injected binary by comparing the waveforms generated from their respective source parameters. For each reported source, we construct a representative parameter estimate using the median of the samples for nonperiodic parameters and the circular mean for periodic parameters. We then generate the corresponding LISA TDI waveform and compare it with each injected waveform. This association accounts for the agreement of the waveforms across the TDI channels, including the effects of sky position, amplitude, and other source parameters. The association procedures and noise-weighted inner product are detailed in Appendix~\ref{app:assoc}.

For both association criteria, we define the recovery fraction as the fraction of injected binaries successfully matched to a reported source. For associated sources, we also compare the inferred parameters with their injected values. The calibration of the predicted uncertainties is assessed separately, without selecting only successfully matched sources. In particular, parameter calibration is evaluated conditional on the injected source count, using parameter samples ordered by frequency.

These association procedures require knowledge of the injected sources and are used only to evaluate performance on simulated data. For observations without known injections, \SFG provides candidate sources, their existence probabilities, and their parameter densities, which can subsequently be used to initialize likelihood-based inference, as discussed in Sec.~\ref{sec:globalfit}.

\section{Results}
\label{sec:results}

We now address the four questions outlined in Sec.~\ref{sec:intro} using simulated LISA observations containing varying numbers of overlapping Galactic binaries. We first assess how reliably \SFG identifies individual sources and determines their number. We then investigate how frequency and sky separation affect the recovery of overlapping binaries, before examining the accuracy of the inferred source parameters and their uncertainties. Finally, we evaluate the computational cost and investigate how the learned parameter densities can support subsequent likelihood-based inference. We discuss the dependence of \SFG's performance on model complexity and the number of slots in Appendix~\ref{app:sensitivity}, while deferring a comprehensive hyperparameter sweep to future work.

\subsection{Source recovery and counting}
\label{sec:counting}
\label{sec:mult}
\label{sec:crowd}

\begin{figure*}[t]
\centering
\fig{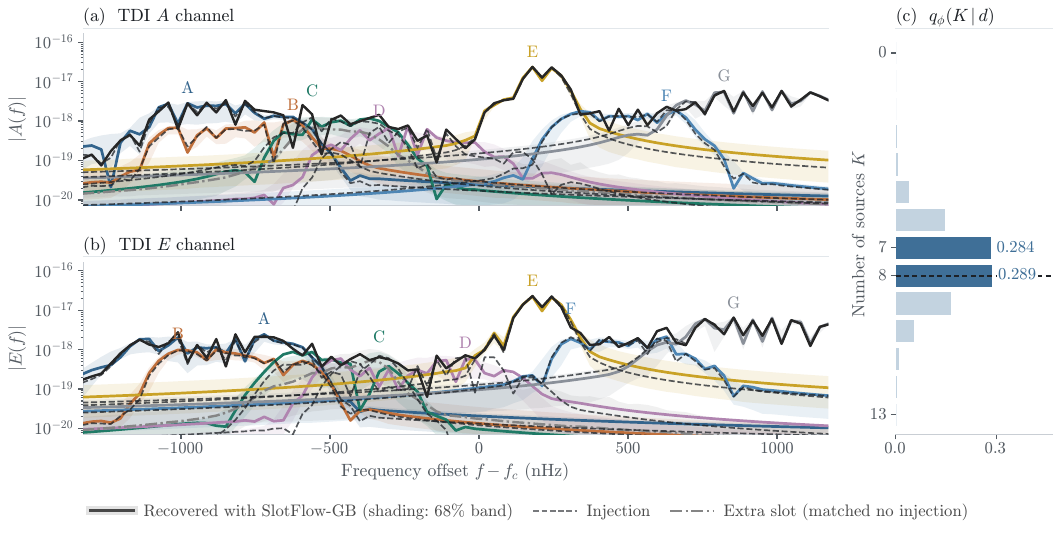}
\caption{
Seven-source inference for the band of Fig.~\ref{fig:task}.
(a) and (b) Reconstructed TDI $A$- and $E$-channel amplitude spectra. For each inferred source, solid curves show the amplitude spectrum evaluated at the median values of its inferred parameters, while shaded bands indicate the pointwise $16$th--$84$th percentile ranges of spectra generated from samples of its parameter distribution. Dashed curves show the injected binaries. Seven of the eight inferred sources correspond to the seven injected binaries. The additional source, shown as a grey dash-dotted curve, lies close to source C and has a broad range of predicted spectral amplitudes, indicating that its contribution is less well constrained.
(c) Inferred source-count distribution $q(K\mid d)$, which assigns comparable probabilities to seven and eight sources. The inferred parameters of Source A are examined further in Fig.~\ref{fig:corner}.
}
\label{fig:showcase}
\end{figure*}

We begin with the seven-source scene introduced in Fig.~\ref{fig:task}, illustrating the simultaneous recovery of overlapping binaries and the inference of their number. Figure~\ref{fig:showcase} shows that \SFG recovers components associated with all seven injected binaries, including sources A and B whose injected frequencies differ by only $73.1\,\mathrm{nHz}$, approximately one Fourier bin of the segmented input. The reconstructed spectra closely follow the injected signals. The network assigns comparable probabilities to seven and eight sources, with a modal count of $\hat K=8$. The additional component lies $19\,\mathrm{nHz}$ from source C in frequency, with a comparable amplitude, and its reconstructed TDI $A$ amplitude spectrum overlaps that of source C by $0.965$, compared with $0.465$ for the next closest injected binary. The eighth component therefore appears to be a duplicate representation of source C rather than a distinct binary.

\begin{figure}[]
\vspace{10pt}
\centering
\includegraphics{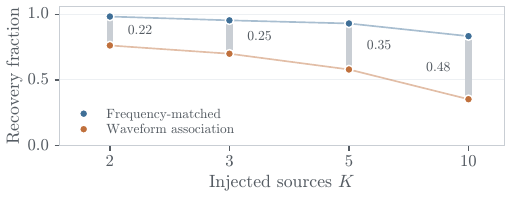}
\caption{
Fraction of injected binaries recovered as a function of source multiplicity on the multiplicity grid evaluation set described in Table~\ref{tab:evalsets}. Frequency matching accepts a source within $500\,\mathrm{nHz}$ of an injected frequency; waveform association requires a normalized TDI waveform match $\mathcal M\geq0.5$.
}
\label{fig:recovery_mult}
\end{figure}

We next examine how source recovery scales with the number of binaries. In the multiplicity grid evaluation set, sources are placed at approximately $200\,\mathrm{nHz}$ intervals, keeping their frequency separation comparable as their number increases. Figure~\ref{fig:recovery_mult} shows that frequency-matched recovery remains high, decreasing only moderately from $0.98$ at $K=2$ to $0.83$ at $K=10$. Waveform-associated recovery decreases more strongly, from $0.76$ to $0.35$. Thus, \SFG continues to locate most binaries in frequency even at high multiplicities, while recovering their full waveforms becomes more challenging.

\begin{figure}[]
\vspace{10pt}
\centering
\fig{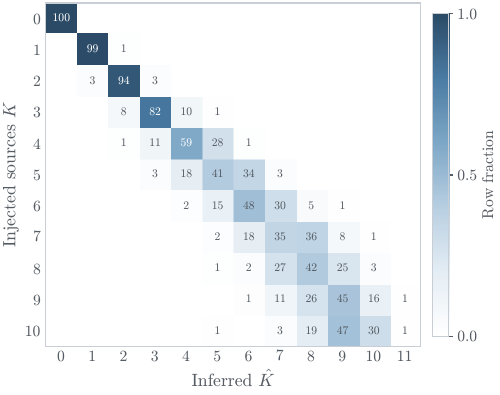}
\caption{
Source-count confusion matrix for the population set (Table~\ref{tab:evalsets}), covering scenes with $K=0$ to $10$ binaries. Each row shows the percentage of scenes assigned a predicted count $\hat K$ for a given true count $K$. The network identifies the exact number of binaries in $61\%$ of scenes, while in $94\%$ of scenes the predicted count is either correct or differs by only one binary.
}
\label{fig:conf}
\end{figure}

We next assess how accurately \SFG determines the number of binaries. Figure~\ref{fig:conf} compares the inferred source count $\hat K$ with the true count $K$ across scenes containing between zero and ten binaries. The network identifies the exact number in $61\%$ of scenes, while in $94\%$ its prediction is either correct or differs by only one binary. Larger counting errors are therefore uncommon, although undercounting becomes more frequent at higher multiplicities. As illustrated by the seven-source example in Fig.~\ref{fig:showcase}, an overcount can arise when multiple slots represent the same binary, suggesting a possible avenue for improving source-count estimates by identifying duplicate components.

In addition to estimating the most likely number of binaries, \SFG provides a probability distribution $q(K\mid d)$ over possible source counts. This distribution allows us to assess not only whether the network predicts the correct number, but also how much probability it assigns to that number relative to other possibilities. For each simulated scene, we evaluate the probability assigned to the injected source count $K_{\mathrm{true}}$ and compute its negative logarithm, $-\ln q(K_{\mathrm{true}}\mid d)$. Averaging over the evaluation set gives a mean score of $1.19$, where lower values indicate better predictions. For comparison, we obtain $2.40$ if we ignore the LISA observations entirely and assign probabilities using only how frequently each source count occurs in the simulated population. The difference of $1.21$ in log probability means that \SFG assigns approximately $e^{1.21}\sim3.4$ times as much probability to the true source count. This demonstrates that the network extracts substantial information about source multiplicity from the observed signals.

\begin{figure}[]
\centering
\fig{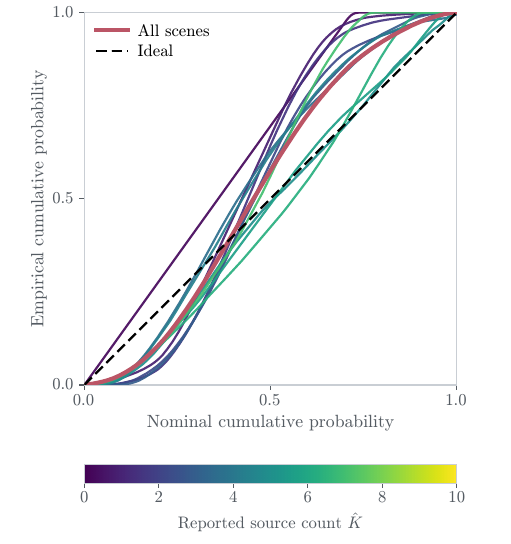}
\caption{
Calibration of the predicted source-count distribution $q(K\mid d)$, assessed using the probability integral transform (PIT). For each scene, the PIT measures where the true number of binaries falls within the predicted cumulative probability distribution. If the predicted probabilities correctly describe the uncertainty, these values are uniformly distributed and the resulting curves follow the dashed diagonal. The thick red curve includes all scenes, while the thinner curves show results separately for each most probable source count $\hat K$, indicated by the color scale. The calculation accounts for the discrete source count and uses the full predicted distribution over $K=0,\ldots,13$. The treatment of discrete source counts and construction of the reference band are detailed in Appendix~\ref{app:sbc}.}
\label{fig:ppcount}
\end{figure}

We next examine whether the probabilities assigned to different source counts accurately reflect the network's uncertainty. Figure~\ref{fig:ppcount} shows a probability--probability (p--p) plot, where correctly calibrated probabilities would follow the diagonal. The pooled curve departs systematically from this reference, showing that \SFG generally assigns more uncertainty to the number of binaries than its observed counting errors would suggest. This is particularly pronounced at low reported multiplicities: when the network predicts one binary, that count is correct in $97.5\%$ of scenes, yet its probability distribution remains considerably broader than this accuracy would suggest. The confusion matrix in Fig.~\ref{fig:conf} also shows that counting becomes more difficult at higher multiplicities, where undercounting is increasingly common.

The construction of $q(K\mid d)$ contributes to the calibration discrepancy. The distribution combines existence probabilities from thirteen source slots and therefore assigns probability to counts up to $K=13$, although the simulated scenes contain at most ten binaries. On average, $3.6\%$ of the predicted probability lies above this range, increasing to approximately $30\%$ when the reported count is $\hat K=10$. Despite this excess uncertainty, the network identifies a small set of plausible source counts with high accuracy. The two most probable counts include the true number of binaries in $83\%$ of scenes, increasing to $93\%$ for the three most probable counts. These predictions provide candidate source counts for subsequent likelihood-based inference, allowing several plausible hypotheses to be explored without committing to a single estimate.

\subsection{Disentangling overlapping binaries}
\label{sec:disentangle}
\label{sec:fsep}
\label{sec:skysep}
\label{sec:surface}
\label{sec:ceiling}
\label{sec:param}

\begin{figure*}[t]
\centering
\fig{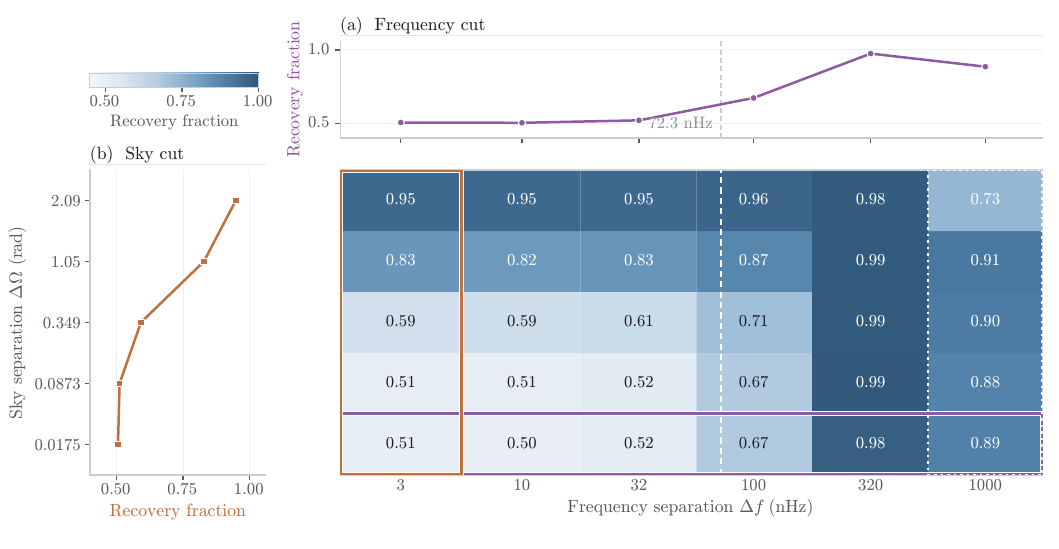}
\caption{
Recovery across the plane of frequency separation $\Delta f$ and angular separation $\Delta\Omega$ for controlled two-source scenes. The heatmap shows the measured per-source recovery fraction; panel (a) is the frequency cut at the smallest angular separation and panel (b) the sky cut at the smallest frequency separation, each outlined on the heatmap in its own color. The dashed vertical line marks the $72.3\,\mathrm{nHz}$ frequency spacing of the segmented representation. The dotted outline at the largest frequency separation marks the recovery decline examined in Appendix~\ref{app:anchor}.
}
\label{fig:res}
\end{figure*}

We next investigate how the separation between two binaries in frequency and sky position affects their recovery by \SFG. In controlled two-source scenes, we vary frequency separation $\Delta f$ and angular separation $\Delta\Omega$ independently. Figure~\ref{fig:res} shows that for binaries close together on the sky, recovery increases as their frequencies move apart. Conversely, at a separation of only $\Delta f=3\,\mathrm{nHz}$, frequency-matched recovery rises from $0.51$ to $0.95$ as the binaries move farther apart on the sky. Waveform association shows the same qualitative dependence on sky separation, providing a complementary assessment of source recovery at separations well below the $500\,\mathrm{nHz}$ frequency-matching tolerance. Recovery decreases again at the largest tested frequency separations, where the current frequency-reference procedure can mislocalize a source. Appendix~\ref{app:anchor} examines this effect and its sensitivity to the matching tolerance. Together, these results show that \SFG can exploit the sky-dependent amplitude and phase modulations of the LISA response to distinguish binaries with nearly identical frequencies.

\begin{figure}[t]
\centering
\fig{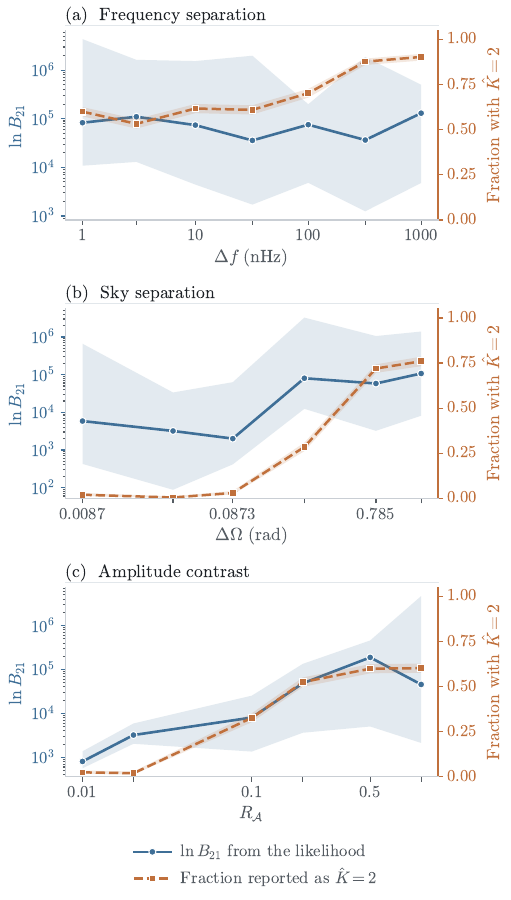}
\caption{
Comparison of the evidence for two binaries with the source-count predictions of \SFG. Blue curves show the approximate log Bayes factor $\ln B_{21}$ in favor of two sources over one, while orange curves show the fraction of scenes for which the network predicts the correct count, $\hat K=2$. Panels (a)--(c) vary frequency
separation, sky separation, and amplitude contrast, respectively.
Each cell contains $300$--$400$ scenes. The blue band spans the $16$th to $84$th percentile of $\ln B_{21}$ across the scenes of a cell, floored where the lower percentile falls below the plotted range; the orange band is the Clopper--Pearson~\cite{CLOPPER1934} $68\%$ interval on the binomial fraction.
}
\label{fig:evidence}
\end{figure}

We next investigate whether the limits in source recovery reflect how much information is available in the LISA data. When \SFG identifies only one binary in a scene containing two, we ask whether the data themselves provide sufficient evidence to distinguish both signals. We therefore compare the Bayesian evidence for models containing one and two binaries, quantified by the Bayes factor $B_{21}=Z_2/Z_1$, where $Z_K$ is the evidence for $K$ sources. Values of $B_{21}>1$ favor the two-source model. We estimate the evidence using a Laplace approximation, which treats the likelihood near its maximum as a multivariate Gaussian and uses its width to estimate the parameter-space volume supported by the data. This is computationally efficient for the high-SNR binaries considered here, whose likelihoods are strongly concentrated around their maxima. For the one-source model, we optimize the parameters of a single binary to reproduce the combined signal from both injected binaries as closely as possible. Since the likelihood can have several local maxima, we repeat this optimization from different initial guesses and use the solution with the highest likelihood. This reduces the risk of favoring the two-source model simply because a good one-source fit was missed. Figure~\ref{fig:evidence} compares the resulting $\ln B_{21}$ with the fraction of scenes in which \SFG identifies two binaries, as we vary their frequency separation, angular separation, and amplitude ratio. Further details and numerical checks are given in Appendix~\ref{app:bayes}.

The approximate Bayesian evidence favors two binaries throughout the high-SNR configurations considered here, with $\ln B_{21}\geq14.7$. However, the network's ability to identify both sources depends strongly on their configuration. In Fig.~\ref{fig:evidence}(a), the fraction of scenes in which the network reports two binaries increases from approximately $0.60$ to $0.90$ as their frequency separation grows, while the evidence remains strongly in favor of two sources across the entire range. The sky-separation experiment in panel (b) shows a similar distinction: although the evidence supports two binaries even at small angular separations, the network identifies both in only $2\%$ of scenes at the smallest separation. In panel (c), increasing the amplitude of the fainter binary improves both the evidence and the network's recovery, but at the largest amplitude contrast the network identifies two sources in only $3\%$ of scenes despite the evidence favoring the two-source model.

These results show that strong likelihood-based evidence for two binaries does not necessarily translate into a correct source-count prediction by \SFG. In several configurations, the approximate evidence strongly favors two sources while the network predicts only one. This identifies limitations of the learned counting procedure that are not solely determined by the distinguishability of the signals in the data.

\subsection{Source parameters and uncertainties}
\label{sec:calib}
\label{sec:reference}

\begin{figure*}[t]
\centering
\includegraphics[width=0.80\textwidth]{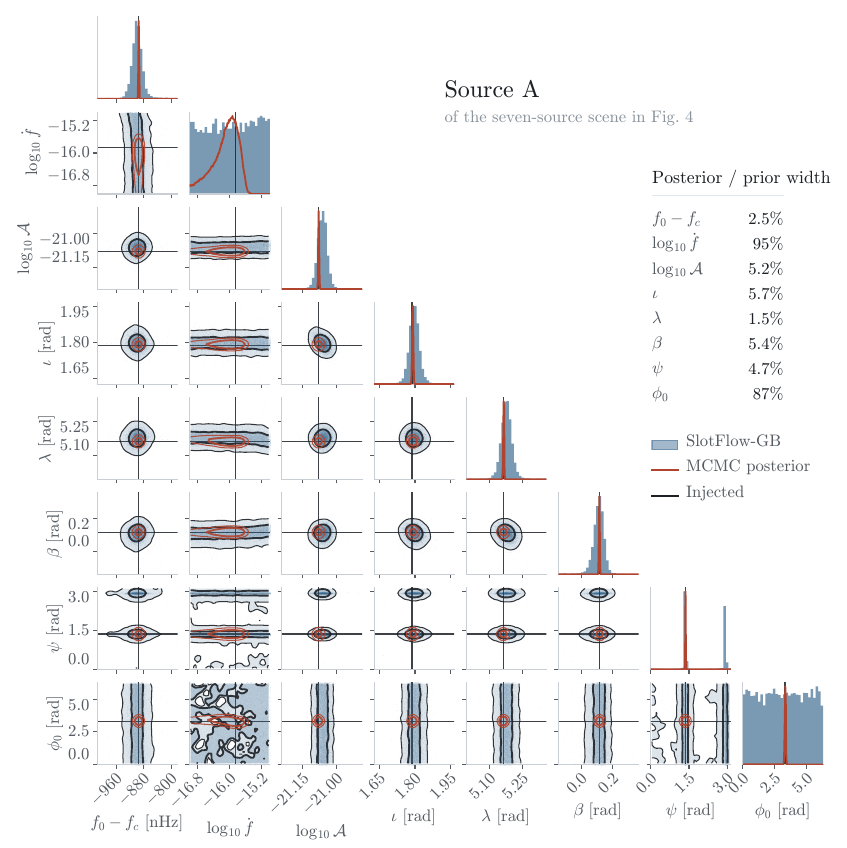}
\caption{
Eight-dimensional learned parameter distribution for Source A in Fig.~\ref{fig:showcase}. Black lines mark the injected parameters, and contours enclose $68\%$ and $95\%$ of the plotted draws. The learned distribution captures the amplitude--inclination degeneracy and the exact polarization--phase symmetry $(\psi,\phi_0)\to(\psi+\pi/2,\phi_0+\pi)$. The table reports the $68\%$ highest-density width of each parameter relative to its simulation prior, with $100\%$ indicating an unconstrained parameter. \textcolor{black}{Red curves and contours show the marginal MCMC posterior for Source A, obtained by jointly sampling all seven binaries with their number fixed to $K=7$.} The $\log_{10}\dot f$ panel displays the simulation-prior range, excluding $0.35\%$ of the learned draws; restricting draws to this range changes the tabulated widths by less than $2\%$.
}
\label{fig:corner}
\end{figure*}

We now examine how accurately \SFG determines the physical parameters of individual binaries and how well its predicted distributions describe their uncertainties. We begin with Source A from the seven-source example in Fig.~\ref{fig:showcase}. This binary belongs to the closest pair in frequency and is approximately eight times fainter in peak spectral magnitude than the dominant source E. \textcolor{black}{Figure~\ref{fig:corner} shows the inferred eight-dimensional parameter distribution for source A, together with the marginal of a seven-source reference posterior. Despite the overlap with neighboring signals, the median inferred frequency differs from the injected value by only $3.1\,\mathrm{nHz}$, and the inferred sky position by $0.9^\circ$. The reference posterior was obtained with the parallel-tempered ensemble sampler \textsc{eryn}~\cite{Karnesis:2023ras}, sampling the seven binaries jointly with their number fixed to $K=7$, under the training priors and using the A and E channels. We ran 128 walkers and three temperatures, initialized at the injected parameters perturbed by Gaussian noise with a standard deviation of $3\times10^{-3}$ of each prior width. We kept $9\,500$ steps after thinning by a factor of ten, and discarded the first quarter as burn-in. Because \SFG performs simultaneous search and parameter estimation, additional comparisons could be drawn by running end-to-end global fit pipelines (e.g.~\cite{glass}); since the required computational budget would be significantly higher, this will be the subject of future work.}

\begin{figure}[tb]
\centering
\fig{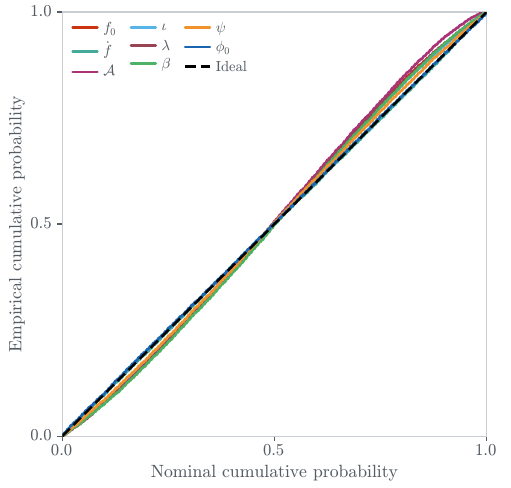}
\caption{
Probability--probability (p--p) plots assessing the calibration of the learned distributions for all eight source parameters. For each observation, parameter samples are drawn from the network's predicted distribution conditional on the injected source count. The binaries are ordered by frequency within each sample, and their parameters are compared with those of the injected binaries in the same frequency order, without matching individual sources. The dashed diagonal represents correctly calibrated uncertainties. All observations containing at least one binary are included, regardless of source-recovery success. The analysis accounts for the joint polarization--phase symmetry. Error bands on these curves are omitted because they are negligible on the scale of the plot. The conditional sampling procedure, frequency ordering, and treatment of periodic parameters are detailed in Appendix~\ref{app:sbc}.
}
\label{fig:ppparam}
\end{figure}

The learned distribution also captures important relationships between the source parameters. In particular, it reproduces the amplitude--inclination degeneracy and both equivalent solutions associated with the exact polarization--phase symmetry $(\psi,\phi_0)\to(\psi+\pi/2,\phi_0+\pi)$. Other parameters are less tightly constrained: the frequency derivative and initial phase retain approximately $95\%$ and $87\%$ of their respective prior widths. The likelihood-based reference is substantially narrower, but it is given information that \SFG has to work out for itself: the true source count, and an initialization at
the injected parameters. It also explores only one of the two symmetry branches, while the network represents both and reports $\hat K=8$ for this scene. The reference is therefore a local characterization of the mode around the injection rather than a converged posterior; the calibration of the learned widths is assessed over 6000 scenes in Fig.~\ref{fig:ppparam}.

Two factors plausibly contribute to the weak constraint on $\dot f$. The frequency derivative imprints only $\pi\dot f T_{\rm obs}^{2}$ of additional phase over the observation, remaining below one radian for about three quarters of the simulation prior. Part of this
signature may also be absorbed by a shift in $f_0$. Separately, the fixed preprocessing transforms eight overlapping segments, each with a frequency resolution of $1/T_{\rm seg}=\SI{72.3}{\nano\hertz}$ rather than the full-baseline $1/T_{\rm obs}=\SI{32.15}{\nano\hertz}$. We have not isolated the contribution of either factor.

The broad $\phi_0$ distribution is likewise consistent with the frequency uncertainty: a frequency uncertainty $\delta f$ corresponds to a phase shift of approximately $\pi\,\delta f\,T_{\rm obs}$. For Source A, the \SI{34}{\nano\hertz} width of the learned frequency distribution corresponds to $3.3$ rad, about half of the $2\pi$ prior, compared with $0.19$ rad for the \SI{2}{\nano\hertz} width of the reference posterior.
The exact polarization--phase symmetry further produces two equivalent solutions separated by $\pi$ in $\phi_0$.

We next assess the parameter accuracy across the evaluation sets. For successfully associated sources in the broadly sampled two-source set, the median absolute frequency error is $6.5\,\mathrm{nHz}$, and the median sky-position error is $1.2^\circ$. The corresponding median reported uncertainties are $10.9\,\mathrm{nHz}$ for the $68\%$ frequency interval and \textcolor{black}{$8.8\,\mathrm{deg}^2$} for the $90\%$ sky region. Parameter estimation becomes more challenging as the number of binaries increases: in the multiplicity evaluation, the median frequency error among associated sources rises from $7.5\,\mathrm{nHz}$ at $K=2$ to $42.2\,\mathrm{nHz}$ at $K=10$, consistent with the reduction in waveform-associated recovery reported in Fig.~\ref{fig:recovery_mult}.

We next examine whether the learned parameter distributions provide calibrated marginal uncertainties when the number of binaries is known. This separates parameter uncertainty from errors in the predicted source count.

Comparing predicted and injected binaries normally requires deciding which recovered source corresponds to which injection. Such matching can introduce selection effects by favoring sources that the network recovers accurately. To avoid this, we evaluate the predicted parameter distributions without matching individual sources. For each observation, we use the known number of injected binaries to select the network's predicted distribution for that source count, even if the network considers a different count more likely. We then draw 256 samples from this distribution, each consisting of a set of GB parameters with the same number of sources as the observation. Within each sample, we sort the binaries by frequency, and do the same for the injected binaries. We can then compare the lowest-frequency injected binary with the predicted distribution for the lowest-frequency binary, the second-lowest with the second-lowest, and so on. For each parameter, we calculate the fraction of samples in which the predicted value falls below the corresponding injected value. If the predicted uncertainties are correctly calibrated, these fractions are uniformly distributed and the p--p curves follow the diagonal.

This procedure does not require assigning individual predictions to injected binaries. Even when closely spaced sources exchange their frequency order between samples, the comparison remains well defined because each sample is sorted independently. The resulting p--p plots assess parameter calibration conditional on the true source count and include all observations containing at least one binary, including cases where the network fails to recover individual sources.

Figure~\ref{fig:ppparam} shows that the p--p curves remain relatively close to the diagonal, although several exhibit systematic deviations indicating mildly conservative uncertainties. For the six parameters meaningfully constrained by the network, nominal $50\%$ intervals contain the injected values in $53$--$56\%$ of cases, while nominal $90\%$ intervals contain them in $92$--$94\%$ of cases. The learned distributions therefore provide reasonably well-calibrated uncertainties for individual source parameters, with a tendency to slightly overestimate their uncertainty. 

An additional challenge arises when the parameters of neighboring binaries are strongly correlated. The learned normalizing flows capture correlations among the eight parameters of each individual source, as illustrated in Fig.~\ref{fig:corner}. However, the factorized distribution of Eq.~\eqref{eq:factorized} does not explicitly represent correlations between different binaries. These correlations can become important when two signals overlap strongly, since changing the parameters of one binary can be partly compensated by changes in the other.

\begin{figure}[tb]
\centering
\fig{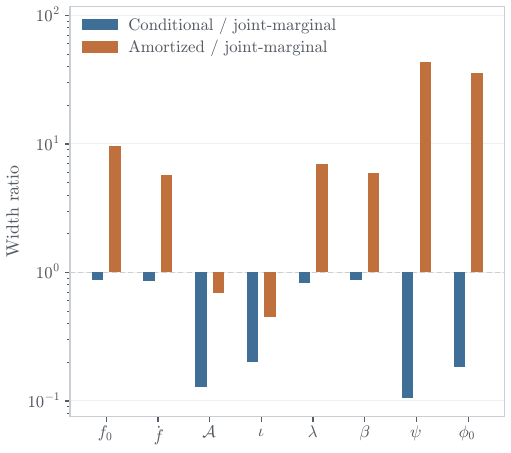}
\caption{
Comparison of parameter uncertainty widths for a strongly overlapping two-binary system. All widths are divided by the joint-marginal Fisher uncertainty obtained when both binaries are allowed to vary. Blue bars show the conditional Fisher uncertainties when the neighboring binary is held fixed; orange bars show the uncertainties predicted by \SFG. Values below one indicate narrower uncertainties than the joint-marginal reference, while values above one indicate broader uncertainties. The dashed horizontal line marks equal widths. The Fisher estimates use a local linearized-signal approximation at the injected parameters; further details are given in Appendix~\ref{app:fisher}.
}
\label{fig:jointk2}
\end{figure}

We illustrate the importance of correlations between overlapping binaries in Fig.~\ref{fig:jointk2}, using a pair separated by only $5.1\,\mathrm{nHz}$ in frequency and $0.0087$ rad on the sky. As a reference, we calculate the parameter uncertainties from the joint Fisher matrix, allowing both binaries to vary and marginalizing over the neighboring source. The figure shows uncertainty widths relative to this reference, with a ratio of one indicating equal widths. The blue points show the effect of fixing the neighboring binary: the resulting uncertainties are smaller, particularly for amplitude, inclination, polarization, and initial phase, where they can be reduced by nearly an order of magnitude. This demonstrates how strongly the inferred parameters of one binary can depend on the uncertainty in another.

The orange points compare the uncertainties predicted by \SFG with the same joint-marginal Fisher reference. The learned distributions are considerably broader for several parameters, particularly frequency, sky position, polarization, and initial phase, while their amplitude and inclination uncertainties are narrower. \nhouba{Polarization and initial phase are a special case: the two symmetry-related solutions differ by $\pi/2$ in $\psi$ and $\pi$ in $\phi_0$, and \SFG represents both while the Fisher estimate describes only one. Its predicted $\psi$ distribution is correspondingly about $\pi/2$ wide, so the large ratios for these two parameters measure the separation between the two solutions rather than the precision with which either is determined.} \nhouba{Overall, the}
learned uncertainties do not consistently reproduce the local likelihood-based estimates, even when correlations between sources are accounted for in the reference. Since the Fisher matrix provides only a local Gaussian approximation, these differences do not necessarily imply incorrect posterior coverage, but illustrate the limitations of interpreting the learned distributions as substitutes for a full joint likelihood analysis.

Overall, \SFG provides informative parameter estimates and reasonably calibrated uncertainties, even for overlapping binaries, while capturing important degeneracies within individual sources. However, the comparison with the Fisher estimates highlights the importance of correlations between neighboring binaries, which are not explicitly represented by the current factorized model. \textcolor{black}{These results characterize both the capabilities and current limitations of the learned posterior approximation, particularly the absence of explicit cross-source correlations.}

\subsection{Computational performance and likelihood refinement}
\label{sec:scaling}
\label{sec:globalfit}

\textcolor{black}{We next examine the computational cost of this amortized representation and its practical use in subsequent likelihood-based inference.} A principal advantage of amortized inference is that, once trained, \SFG predicts the source count and parameter distributions without performing a separate likelihood search for each observation. On an NVIDIA GH200 GPU, inference takes $41\,\mathrm{ms}$ per frequency band when processing one observation at a time, decreasing to $14\,\mathrm{ms}$ per band when processing batches of $32$. These timings include drawing $1000$ samples from the predicted parameter distributions. On an Apple M2 Max CPU using eight threads, a single-band evaluation takes $0.53\,\mathrm{s}$. Training required approximately $112$ GPU-hours across the eight GH200 GPUs. Thus, the computational cost is concentrated in training, which happens once based on the simulation domain, while subsequent observations can be analyzed at low per-band cost.

In this subsection, we investigate whether \SFG's predicted distributions can accelerate likelihood-based parameter estimation. We return to Source A in the seven-binary example of Fig.~\ref{fig:showcase}, and refine its eight parameters using a likelihood-based sampler, with the other six binaries fixed at their injected values. The sampler, described in detail in Appendix~\ref{app:handoff}, is run with identical settings and computational budgets, but different initializations: samples from the distribution learned by \SFG, samples from the simulation prior, samples from the injected parameters themselves, and samples from a Fisher approximation centered on the injection. The last two provide favorable reference cases that require knowledge of the true parameters and are therefore unavailable in a blind search.

\begin{figure}[tb]
\centering
\fig{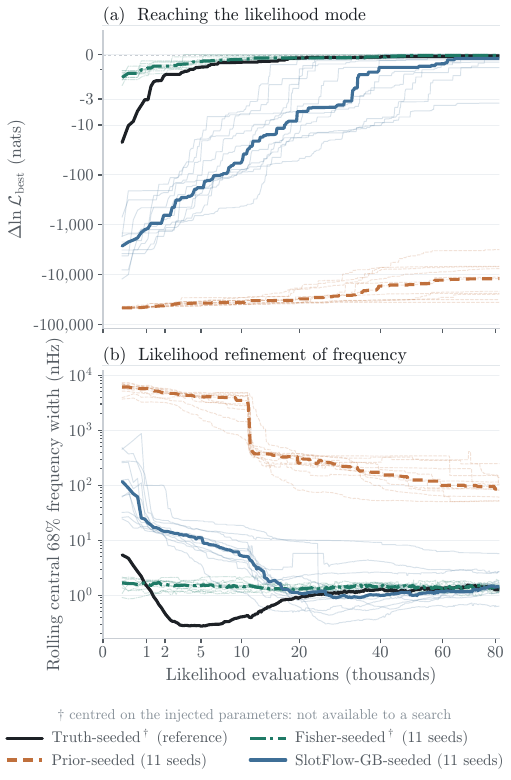}
\caption{
Likelihood-based refinement of Source A in Fig.~\ref{fig:showcase}, with the other six binaries fixed at their injected parameters. The sampler is initialized using the \SFG distribution, the simulation prior, the injected parameters, or a Fisher approximation centered on the injection. The latter two require knowledge of the true parameters and serve as reference cases. Each initialization is tested with $11$ independent runs. Panel (a) shows the best log likelihood reached relative to the highest value obtained from truth-initialized runs. Panel (b) shows the central $68\%$ frequency width of the selected high-likelihood walkers. Thick lines indicate medians across runs, and faint lines show individual runs. The horizontal axis uses a square-root scale.  The sampler configuration, likelihood threshold, and frequency-width calculation are detailed in Appendix~\ref{app:handoff}.
}
\label{fig:handoff}
\end{figure}

Figure~\ref{fig:handoff} shows how the best likelihood and the estimated frequency uncertainty evolve with the number of likelihood evaluations. The learned distribution provides a substantially better starting point than the simulation prior. As shown in Fig.~\ref{fig:handoff}(a), the sampler reaches within one nat of the best likelihood obtained from truth-initialized runs in $10$ of $11$ runs started from the learned distribution. None of the prior-initialized runs reaches this threshold within the same computational budget. Figure~\ref{fig:handoff}(b) shows that the selected high-likelihood walkers in the learned-density runs also reach nanohertz-scale frequency widths comparable to the truth-centered references, although they require more likelihood evaluations. The network therefore provides initial parameter configurations from which a likelihood sampler can efficiently approach a high-likelihood solution, without access to the injected parameters.

Importantly, the advantage comes from using the learned distribution, not simply its central parameter estimate. Initializing a narrow Gaussian ensemble around the median of the network's predictions performs substantially worse in this example. In particular, the initial phase remains weakly constrained, so its median does not necessarily correspond to a favorable likelihood solution. Drawing initial configurations from the full learned distribution retains this uncertainty and increases the chance of starting in a region of high likelihood. These results demonstrate that the learned parameter distributions retain information useful for likelihood-based refinement, providing a direct connection between amortized probabilistic inference and conventional parameter estimation. The present experiment establishes this benefit for conditional single-source refinement; extending it to joint inference over multiple binaries and an unknown source count remains future work.

\section{Discussion and outlook}
\label{sec:remains}
\label{sec:conclusions}

We have investigated several aspects of amortized inference for overlapping Galactic binaries in LISA data, including the recovery and counting of sources, the separation of closely overlapping signals, the accuracy of inferred parameters and uncertainties, as well as \textcolor{black}{the computational performance of the learned model and its utility for subsequent likelihood-based analysis.} Our results demonstrate that \SFG can identify multiple binaries, estimate their number, and infer individual source-parameter distributions in a single network evaluation, thus representing an efficient end-to-end framework for simultaneous GB search and amortized parameter estimation. The controlled experiments further show that the network can exploit the sky-dependent modulation of the LISA response to distinguish binaries even when their frequencies are nearly identical. At the same time, comparisons with approximate Bayesian evidence reveal configurations in which the data strongly favor two sources but the network identifies only one. This indicates that the present recovery limits are not always imposed by the information available in the observations, leaving room for improving the learned source-discovery procedure.

Beyond identifying sources, \SFG provides informative parameter distributions that capture important within-source degeneracies and yield reasonably calibrated uncertainties when evaluated conditional on the true source count. However, the source-count probabilities are not yet fully calibrated, and the factorized representation does not explicitly capture correlations between different binaries. The latter can become particularly important for strongly overlapping signals, as illustrated by the Fisher comparison. \textcolor{black}{These limitations motivate further development of the probabilistic approximation, particularly toward joint inference over source multiplicity and correlated source parameters.} \textcolor{black}{The conditional refinement experiment separately demonstrates the practical utility of the learned distributions: samples drawn from them allow a likelihood sampler to reach high-likelihood solutions substantially more reliably than initialization from the simulation prior, with the network itself requiring only milliseconds per frequency band after training to produce them.}

The present study deliberately focuses on a controlled, relatively high-SNR regime to isolate the challenges of overlapping-source inference. The simulations cover restricted frequency and amplitude ranges, at most ten binaries per observation, stationary Gaussian instrumental noise, and no unresolved Galactic foreground. Consequently, the results do not yet establish performance for the weaker and more numerous binaries expected in realistic LISA observations. Extending the approach to lower-SNR sources, larger source populations, realistic Galactic foregrounds, and uncertain or time-dependent instrumental noise is therefore essential. Further challenges include robustness to gaps and glitches, improved source-count calibration, more effective allocation of signal information across source slots, and the representation of cross-source uncertainties.

\textcolor{black}{Taken together, these results establish a proof of principle for neural trans-dimensional inference of overlapping Galactic binaries, combining source-count estimation with probabilistic characterization of individual sources in a single amortized evaluation.} \textcolor{black}{The demonstrated ability to initialize likelihood-based inference provides a pathway toward integrating these learned distributions into a joint, trans-dimensional analysis.} \textcolor{black}{An important next step is to determine whether this integration yields an end-to-end computational advantage in a realistic LISA global fit.} \textcolor{black}{More broadly, the framework illustrates how learned probabilistic representations of variable source populations can complement likelihood-based methods in complex, overlapping-signal inference problems.}

\begin{acknowledgments}
This research was funded by the Gravitational Physics Professorship at ETH Zurich and supported by the SIGMA-GW grant of the Swiss AI Initiative (project a0188). The authors thank Michele Vallisneri for insightful discussions and guidance on the scientific interpretation and presentation of the results, and Arianna Renzini for valuable discussions on LISA data analysis and catalog construction, as well as for feedback about the manuscript. The authors also thank the LISA Simulation Working Group and the LISA Simulation Expert Group for stimulating discussions on LISA simulation and data-analysis activities.

The computations were performed on the Alps supercomputer at the Swiss National Supercomputing Center (CSCS) under project a0188 and go122. The authors gratefully acknowledge the support of EuroHack26 at CSCS and, in particular, William Sawyer and Jean-Guillaume Piccinali for their mentorship and advice on high-performance computing and the effective use of the CSCS infrastructure.

OpenAI Codex (GPT-6) and Anthropic Claude (Opus 5) were used to assist with manuscript restructuring, consistency checking of scientific claims, and the development and refinement of plotting, evaluation, and validation code. All numerical results reported here derive from the underlying simulation and evaluation outputs, and AI-assisted text, code, and analyses were reviewed by the authors.
\end{acknowledgments}

\section*{Data Availability}

The software, trained model, simulation and evaluation configurations, and numerical data underlying the figures and results of this work are available from the corresponding author upon reasonable request. 

\appendix

\section{\SFG architecture}
\label{app:arch}

\paragraph{Input and segmentation.}
The simulated TDI time series are sampled at a cadence of $59.3\,\mathrm{s}$. We divide the observations into eight Hann-windowed segments of $233\,012$ samples ($160$ days), separated by a hop of $41\,609$ samples ($28.6$ days), so that the eight segments together span the full observation with approximately $82\%$ overlap. The resulting segment spectra have a frequency spacing of $72.3\,\mathrm{nHz}$.

\paragraph{Local spectral representation.}
The local spectral window is centered on the strongest peak in the combined $A$- and $E$-channel amplitudes, summed across all eight time segments and weighted by a Gaussian centered on the nominal frequency $f_c$. Around this peak, we select $64$ frequency bins (approximately $4.6\,\mathrm{\mu Hz}$) and extract the same frequency interval from both channels in every segment. This allows the network to observe the temporal evolution of the signal amplitudes and phases within a common frequency band.

A convolutional encoder processes the $A$ and $E$ spectra as a ten-channel time--frequency representation. The input channels comprise the log amplitudes of $A$ and $E$, the sine and cosine of each channel's phase, the normalized real and imaginary parts of $AE^{*}$, the log-amplitude ratio, and the $A$--$E$ cross phase. Four $3\times3$ convolutional blocks with group normalization and Gaussian error linear unit (\textsc{gelu}; \cite{gelu}) activations map these inputs to $32$, $64$, $128$, and $128$ channels. Downsampling acts only along frequency, preserving the eight time segments. The encoder produces $512$ time--frequency features of dimension $128$.

In addition, we compute $21$ summary features for each time segment, using the $A$, $E$, and $T$ spectra within the selected frequency window. These features characterize spectral power, relative channel responses, spectral structure, and phase evolution. Their definitions are given in Table~\ref{tab:summaryfeatures}. A two-layer network ($21\to64\to32$ with \textsc{gelu} activations) projects the summaries to $32$ dimensions, which are appended to the corresponding convolutional features. Each time--frequency feature therefore has dimension $160$. The convolutional encoder operates on $A$ and $E$, while information from the $T$ channel enters through the summary features. Learned embeddings identifying the time segment and frequency-bin position are added to the feature vectors before they are passed to the source slots.

\begin{table*}[t]
\caption{
The $21$ summary features computed for each time segment. The features are calculated from the complex Fourier spectra of the $A$, $E$, and $T$ channels within the selected $64$-bin frequency window. Here, $b$ labels a frequency bin, and $A(b)$, $E(b)$, and $T(b)$ denote the corresponding Fourier coefficients. The total spectral power in channel $X$ is $P_X=\sum_b|X(b)|^2$, with $I=P_A+P_E+P_T$. The peak bin $b^\ast$ is the frequency bin where the combined amplitude $|A(b)|+|E(b)|$ is largest, while $\bar b$ denotes the mean frequency-bin position weighted by this combined amplitude. Quantities with a subscript $\ast$ are evaluated at the peak bin, and $\hat X_\ast=X(b^\ast)/|X(b^\ast)|$ denotes its complex Fourier coefficient normalized to unit magnitude. Finally, $n=64$ is the number of frequency bins, and $t_{\rm seg}$ denotes the time associated with the segment.
}
\label{tab:summaryfeatures}
\footnotesize
\begin{ruledtabular}
\begin{tabular}{rll}
\# & Feature & Definition \\
\hline
1 & \texttt{log\_power\_A} & $\ln P_A$ \\
2 & \texttt{log\_power\_E} & $\ln P_E$ \\
3 & \texttt{log\_power\_T} & $\ln P_T$ \\
4 & \texttt{ae\_ratio} & $\tanh(\ln P_A-\ln P_E)$ \\
5 & \texttt{te\_ratio} & $\tanh(\ln P_T-\ln P_E)$ \\
\hline
6 & \texttt{stokes\_Q} & $(P_A-P_E)/I$ \\
7 & \texttt{stokes\_U} & $(P_E-P_T)/I$ \\
8 & \texttt{stokes\_U\_real} & $2\,\mathrm{Re}\sum_b A(b)E^\ast(b)\big/\sum_b\left(|A(b)|^2+|E(b)|^2\right)$ \\
\hline
9 & \texttt{ae\_corr} & $\mathrm{Re}\left[\hat A_\ast\hat E_\ast^\ast\right]$ \\
10 & \texttt{sin\_ae} & $\sin\arg\left[\hat A_\ast\hat E_\ast^\ast\right]$ \\
11 & \texttt{cos\_ae} & $\cos\arg\left[\hat A_\ast\hat E_\ast^\ast\right]$ \\
\hline
12 & \texttt{peak\_idx} & $b^\ast/(n-1)$ \\
13 & \texttt{centroid} & $\bar b/(n-1)$ \\
14 & \texttt{et\_phase} & $\arg\left[E(b^\ast)T^\ast(b^\ast)\right]$ \\
\hline
15 & \texttt{A\_phase\_pk} & $\arg A(b^\ast)$ \\
16 & \texttt{E\_phase\_pk} & $\arg E(b^\ast)$ \\
17 & \texttt{A\_phase\_grad} & $\arg\left[A(b^\ast+1)A^\ast(b^\ast-1)\right]$ \\
18 & \texttt{E\_phase\_grad} & $\arg\left[E(b^\ast+1)E^\ast(b^\ast-1)\right]$ \\
19 & \texttt{width} & $\left[\sum_b(b-\bar b)^2S(b)\big/\sum_b S(b)\right]^{1/2}/(n-1)$ \\
\hline
20 & \texttt{sin\_detrend\_A} & $\sin\left(\arg A(b^\ast)-2\pi f_c t_{\rm seg}\right)$ \\
21 & \texttt{cos\_detrend\_A} & $\cos\left(\arg A(b^\ast)-2\pi f_c t_{\rm seg}\right)$ \\
\end{tabular}
\end{ruledtabular}
\end{table*}

\paragraph{Competitive source slots.}
The model uses $K_{\max}=13$ source slots, each represented by a $160$-dimensional vector. In each slot-attention iteration~\cite{slots}, the slots provide queries and the encoded spectral features provide keys and values. Attention weights are normalized across slots for each feature, allowing the slots to compete for the available signal information. Each slot aggregates its assigned features and updates its representation through a gated recurrent unit~\cite{cho2014learningphraserepresentationsusing} followed by a residual multilayer perceptron. This procedure is repeated for three iterations.

Each final slot representation is passed to four small prediction networks, consisting of a linear layer, a \textsc{gelu} activation, and a second linear layer. These predict the source-existence probability $p_i$, the sky direction, a coarse frequency estimate, and a fine frequency correction. The existence probability is obtained by applying a sigmoid to the corresponding network output.

\paragraph{Local frequency inference.}
Frequency estimation uses a slot-specific spectral reference. Each slot first predicts an approximate source frequency, which is used to identify a nearby peak in the observed spectrum. The frequency of this peak defines the anchor $f_i^{\rm a}$, providing a reference from which the network estimates the source's frequency offset. This allows the network to predict a relatively small frequency correction instead of estimating the absolute source frequency directly. For a training slot $i$ matched to injected source $\sigma(i)$, the normalized frequency offset is
\begin{equation}
r_i=\frac{f_{0,\sigma(i)}-f_i^{\rm a}}{F} \ ,
\end{equation}
where $f_{0,\sigma(i)}$ is the injected source frequency, and $F$ is as in Eq.~(\ref{eq:F}). A regression head predicts the offset $\hat f_i$, while the conditional normalizing flow models the remaining residual $\delta_i=r_i-\hat f_i$. At inference, the absolute source frequency is reconstructed as $f_i=f_i^{\rm a}+F(\hat f_i+\delta_i)$.

\paragraph{Conditional density estimator.}
Each source slot conditions a shared normalizing flow that models the eight-dimensional parameter
distribution of the corresponding binary. The flow consists of eight rational-quadratic-spline
coupling layers~\cite{splines} with $24$ bins per transformation, each conditioned by a network
with two hidden layers of $256$ units. The complete network has $2\,145\,828$ trainable
parameters, approximately $1.45$ million of which belong to the conditional flow.

\paragraph{Training objective.}
During training, predicted slots are associated with injected binaries through a one-to-one Hungarian assignment that minimizes angular separation between predicted and injected sky directions. Let $\sigma(i)$ denote the injected source assigned to slot $i$, and let $\mathcal C$ be the set of matched slots, with $N_{\mathcal C}=|\mathcal C|$. We denote the existence probability of slot $i$ by $p_i$, its predicted sky unit vector by $\hat n_i$, its normalized frequency-offset prediction by $\hat f_i$, and its latent representation by $s_i$.

The total training loss combines six terms that supervise source identification, parameter estimation, and the separation of source representations:
\begin{equation}
\begin{split}
\mathcal L ={}&
\mathcal L_{\rm nll}
+3\,\mathcal L_{\rm exist}
+30\,\mathcal L_{\rm fine}\\
&+10\,\mathcal L_{\rm coarse}
+\mathcal L_{\rm sky}
+2\,\mathcal L_{\rm div},
\end{split}
\label{eq:loss}
\end{equation}
where the coefficients balance the numerical contributions of the individual terms. These are defined as
\begin{align}
\mathcal L_{\rm nll}
&=-\frac{1}{N_{\mathcal C}}
\sum_{i\in\mathcal C}
\log q_i\!\left(
\Delta\theta_{\sigma(i)}\mid d
\right),\\
\mathcal L_{\rm exist}
&=\frac{1}{K_{\max}}
\sum_{i=1}^{K_{\max}}
\mathrm{BCE}\!\left(
p_i,\mathbf{1}[i\in\mathcal C]
\right),\\
\mathcal L_{\rm fine}
&=\frac{1}{N_{\mathcal C}}
\sum_{i\in\mathcal C}
\left(\hat f_i-r_i\right)^2,\\
\mathcal L_{\rm coarse}
&=\frac{1}{N_{\mathcal C}}
\sum_{i\in\mathcal C}
H\!\left(
\frac{f_{0,i}-f_{0,\sigma(i)}}{\Delta f_{\rm c}}
\right),\\
\mathcal L_{\rm sky}
&=\frac{1}{N_{\mathcal C}}
\sum_{i\in\mathcal C}
\left(1-\hat n_i\cdot n_{\sigma(i)}\right),\\
\mathcal L_{\rm div}
&=\frac{1}{|\mathcal P|}
\sum_{(k,j)\in\mathcal P}
\left|
\frac{s_k}{\lVert s_k\rVert}
\cdot
\frac{s_j}{\lVert s_j\rVert}
\right|,\\
\mathcal P
&=\{(k,j):k\notin\mathcal C,\ j\in\mathcal C\}.\nonumber
\end{align}

Each term serves a distinct purpose. The negative log-likelihood $\mathcal L_{\rm nll}$ trains the conditional normalizing flow to assign high probability density to the injected source parameters. Here, $q_i$ is the density predicted for slot $i$, and $\Delta\theta_{\sigma(i)}$ is the transformed parameter vector of its assigned source. Its frequency coordinate is the residual $r_i-\hat f_i$, representing the difference between the injected frequency offset and the offset predicted by the regression head. The remaining seven coordinates describe the other transformed source parameters. The prediction $\hat f_i$ is detached when constructing the target, preventing gradients from the density loss from updating the frequency-regression head.

The existence loss $\mathcal L_{\rm exist}$ trains each slot to predict whether it represents an injected binary. It uses binary cross-entropy ($\mathrm{BCE}$), where slots matched to an injected source are assigned a target of $1$ and unmatched slots a target of $0$. We apply label smoothing of $0.05$, using target probabilities of $0.975$ and $0.025$, respectively, to discourage excessively confident predictions during training.

Frequency estimation is trained in two stages. First, the coarse-frequency loss $\mathcal L_{\rm coarse}$ trains each slot to predict approximately where its assigned binary lies in the spectrum. This prediction guides the search for a nearby spectral peak, which serves as the frequency anchor for the subsequent refinement. Because the coarse prediction can initially be far from the correct frequency, we use the Huber loss to prevent large errors from dominating training. For a normalized frequency error $x$, it is defined as
\begin{equation}
H(x)=
\begin{cases}
x^2/2, & |x|\leq 1,\\
|x|-1/2, & |x|>1.
\end{cases}
\end{equation}
Here, $x=(f_{0,i}-f_{0,\sigma(i)})/\Delta f_{\rm c}$ is the difference between the coarse prediction and injected frequency, expressed in units of the frequency scale $\Delta f_{\rm c}$. Small errors are penalized quadratically to encourage accurate localization, while large errors receive a linearly increasing penalty, reducing their influence on the optimization. Once a spectral anchor has been selected, the fine-frequency loss $\mathcal L_{\rm fine}$ trains the regression head to predict the remaining offset between that anchor and the injected frequency. The conditional flow then models the uncertainty in the residual frequency error.

The sky-position loss $\mathcal L_{\rm sky}$ trains each matched slot to predict the direction of its assigned binary. It penalizes disagreement between the predicted sky unit vector $\hat n_i$ and the injected direction $n_{\sigma(i)}$, reaching zero when the two directions coincide.

Finally, the diversity loss $\mathcal L_{\rm div}$ encourages distinct representations for occupied and unoccupied slots. It penalizes similarity between their normalized latent vectors $s_i$, using the set $\mathcal P$ of matched--unmatched slot pairs. This discourages redundant source representations, although it does not explicitly enforce distinct predictions among all occupied slots.

The existence loss is evaluated for all slots, whereas the source-density, frequency, and sky-position losses apply only to slots matched to injected binaries. For observations containing no injected binaries, the matched-source contributions are defined as zero, while the existence loss remains active. The diversity contribution is likewise zero when no matched--unmatched slot pairs are available.

\section{Post-inference source association} \label{app:assoc}
For each observation, recovery metrics use the $\hat K$ slots with the largest existence probabilities, where $\hat K$ is the most probable source count.

We use two complementary association rules. \emph{Frequency matching} measures coarse
localization: recovered components and injections are paired through a one-to-one
Hungarian assignment~\cite{kuhn1955hungarian} in frequency, and a pair with injection frequency $f_i$ is accepted
when
\begin{equation}
|\hat f_j-f_i|\leq500\,\mathrm{nHz},
\end{equation}
where $\hat f_j$ is the frequency of the highest-density sampled parameter vector for recovered component $j$, as used in the main-text frequency-association procedure.

\emph{Waveform association} instead asks whether a recovered component corresponds to the same binary in the full LISA response. Let $\hat h_j$ be the waveform corresponding to the posterior-median parameters of declared component $j$, with circular means used for periodic coordinates. For an injected binary $i$ with waveform $h_i$, we define
\begin{align}
\mathcal M_{ij}
&=
\frac{\bigl|\langle h_i,\hat h_j\rangle\bigr|}
{\sqrt{\langle h_i,h_i\rangle
       \langle \hat h_j,\hat h_j\rangle}},
\label{eq:match}\\[2pt]
\langle a,b\rangle
&=
\sum_{c\in\{A,E,T\}}
\sum_{f\in\mathcal B}
\frac{a_c(f)b_c^{*}(f)}
     {\sigma_c(f)^2},
\label{eq:inner}
\end{align}
where $\mathcal B$ is the \SI{128}{\micro\hertz} waveform band around \SI{3}{\milli\hertz}, containing $3\,983$ Fourier bins, and
$\sigma_c(f)^2=S_c(f)/(4\delta f_{\rm bin})$ is the variance of the real or imaginary part of the simulated Fourier-domain noise in channel $c$, with $S_c(f)$ the corresponding one-sided instrumental-noise power spectral density. Equation~\eqref{eq:inner} defines the complex
noise-weighted overlap, whose real part is the standard noise-weighted inner product~\cite{cutlerflanagan}. Taking its modulus in Eq.~\eqref{eq:match} maximizes the normalized match over a constant relative phase~\cite{owen_match}.

The match $\mathcal M_{ij}$ compares an injected source with a recovered component.
Taking the modulus maximizes over a constant relative phase, while the normalization
removes the overall amplitude scale. A one-to-one Hungarian assignment maximizes the
total waveform match, and an assigned pair is counted as recovered when
\begin{equation}
\mathcal M_{ij}\geq0.5.
\end{equation}
Unassigned injections and assigned pairs below threshold are counted as misses;
unmatched declared components are spurious.

Raising the waveform-match threshold from $0.3$ to $0.9$ reduces waveform-associated recovery, as expected, while leaving frequency-matched recovery unchanged. The sensitivity to the frequency-matching tolerance is examined for a representative case in Appendix~\ref{app:anchor}.

\section{Sensitivity to representation and model capacity}\label{app:sensitivity}

Among the architectural choices tested, changing the segmentation produces the
clearest performance difference. At a matched training epoch, both the source log likelihood and the exact-count accuracy are highest for segment overlaps
between approximately $71\%$ and $82\%$. At $50\%$ overlap, the source log
likelihood decreases by $0.83$ nats and exact-count accuracy by $6.7$\% relative to the best configuration; at $89\%$ overlap, the
corresponding changes are $1.38$ nats and $3.5$\%.

Increasing the complexity of the network or the size of its spectral input does not improve performance within the tested configurations. Enlarging the conditional flow by a factor of $2.7$ in parameter count, doubling the number of spline bins, increasing the number of slot-attention iterations, or widening the local spectral window from $64$ to $128$ bins either leaves performance within the observed run-to-run variation
($\pm0.15$ nats in source log likelihood and $\pm2\%$ in exact-count
accuracy) or makes it worse. These tests therefore point to the representation
presented to the slots, rather than insufficient capacity of the density estimator,
as the more important limitation among the configurations examined.

The number of slots is also a tradeoff. At a matched training epoch, increasing the pool from $K_{\max}=13$ to
$16$ and $20$ slots improves the source log likelihood monotonically, by
$0.26$ and $0.58$ nats, while exact-count accuracy falls monotonically from
$0.507$ to $0.476$ and $0.455$. Additional slots give the one-to-one assignment
more candidates per source, but every additional slot also enters the
Poisson binomial of Eq.~\eqref{eq:poissonbinomial}, broadening the inferred count
distribution. Ranking on source log likelihood alone would select the largest
pool and lose approximately five points of exact-count accuracy, which is the
most affected metric. We therefore use $K_{\max}=13$, accepting a
slightly worse source log likelihood for the better source count.

We also trained two networks with smaller slot budgets, $3$ and $5$ slots, covering source counts up to three and five, against the $13$ slots of the fiducial model. Compared on the scenes each smaller model was trained for, with the $K=13$ model restricted to the same output range, the differences in exact-count accuracy and recovery are comparable to the observed run-to-run spread  of $\pm0.02$ in exact-count
accuracy. The smaller models also give consistently larger median frequency errors, $8.2$ and $7.0\,\mathrm{nHz}$ against $6.5\,\mathrm{nHz}$ for the $K=13$ model. We therefore find no evidence that replacing the single general model with separate multiplicity-specific networks improves source recovery.

The finite slot pool limits the number of binaries that can be represented within a single analyzed band, but larger frequency ranges can be covered by applying the same network to additional bands without retraining, provided their signal characteristics remain within the training distribution. Extending the method to previously unseen frequency regions or realistic unresolved Galactic foregrounds may require further training, potentially through fine-tuning the existing model on representative simulations. Whether such adaptation is sufficient remains to be investigated.

\section{Calibration of source-count and parameter distributions}
\label{app:sbc}

The p--p plots in Figs.~\ref{fig:ppcount} and \ref{fig:ppparam} assess the calibration of the predicted source-count and parameter distributions, respectively. Both use the unmodified \SFG predictions without selecting successfully recovered sources. Parameter calibration is evaluated conditional on the injected source count.

\paragraph{Source-count calibration.}
The source-count distribution $q(K\mid d)$ is obtained from the slot-existence probabilities through Eq.~\eqref{eq:poissonbinomial}. Since the source count $K$ is discrete, its cumulative distribution
has jumps, and the standard probability integral transform (PIT) is not uniformly distributed even for a perfectly calibrated model. We therefore use the randomized PIT~\cite{Czado2009},
\begin{equation}
u=F_d(K_{\rm true}-1)+v\,q(K_{\rm true}\mid d),
\qquad v\sim\mathcal U(0,1),
\label{eq:countpit}
\end{equation}
where $F_d(k)=\sum_{j=0}^{k}q(j\mid d)$ and $F_d(-1)=0$. Here, $v$ places the PIT value uniformly within the cumulative-probability interval assigned to the true source count. Rather than drawing random
values of $v$, we average over this interval analytically when constructing the p--p curves, avoiding additional sampling noise. If the predicted count distributions are calibrated, the resulting PIT values follow a uniform distribution on $[0,1]$, and the p--p curve follows the diagonal.

We retain the full predicted distribution over $K=0,\ldots,13$, including probability assigned to counts above the maximum of ten binaries in the simulated observations. Figure~\ref{fig:ppcount} shows the pooled calibration curve and curves grouped by the most probable predicted count $\hat K$. The shaded $95\%$ reference band is obtained by repeatedly sampling source counts from the predicted distributions and recomputing the pooled p--p curve.

\paragraph{Parameter calibration.}
For each observation, we evaluate the predicted parameter distribution conditional on its injected source count, even when the network assigns a higher probability to another count. We sample subsets of occupied slots from the independent Bernoulli existence probabilities, conditional on selecting exactly $K_{\rm true}$ slots, and draw an eight-dimensional parameter vector from the flow of each selected slot. The binaries in each sample and in the injected observation are then ordered by frequency.

For each physical parameter and frequency-ordered position, we calculate the fraction of posterior samples below the corresponding injected value. These fractions form the p--p plots in Fig.~\ref{fig:ppparam}. The procedure avoids assigning individual network slots to injected sources and includes all observations containing at least one binary. 

Periodic parameters are treated using consistent circular conventions. In particular, the exact polarization--phase symmetry $(\psi,\phi_0)\mapsto(\psi+\pi/2,\phi_0+\pi)$ is handled jointly by mapping both injected values and posterior samples to the same symmetry-equivalent representation. The resulting curves assess marginal parameter calibration at frequency-ordered positions for a known source count, rather than calibration of the full trans-dimensional posterior.

\section{Frequency-anchor limitations}
\label{app:anchor}

Figure~\ref{fig:res} shows an unexpected decrease in source recovery at the largest tested frequency separations, although more widely separated binaries should generally be easier to distinguish. Interestingly, the network still predicts the correct number of binaries in approximately $90\%$ of these scenes. This suggests that the network often recognizes both sources but does not always assign them the correct frequencies.

\begin{figure}[tb]
\centering
\fig{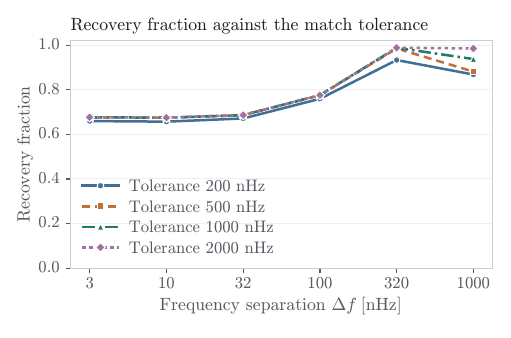}
\caption{
Effect of the frequency-matching tolerance on source recovery in the two-source experiments of Fig.~\ref{fig:res}. A recovered binary is counted as frequency-matched when its estimated frequency differs from an injected frequency by less than the specified tolerance. Results are shown for $200$, $500$, $1000$, and $2000\,\mathrm{nHz}$, with $500\,\mathrm{nHz}$ used in the main analysis. The strong improvement at large frequency separations when the tolerance is relaxed indicates that the decrease in recovery is primarily associated with frequency-localization errors.
}
\label{fig:tolerance}
\end{figure}

To investigate this, Fig.~\ref{fig:tolerance} shows how recovery changes when we vary the maximum frequency difference allowed between an injected and recovered binary. The main analysis uses a tolerance of $500\,\mathrm{nHz}$. Increasing this tolerance substantially improves the measured recovery at large frequency separations, while having little effect at smaller separations. This indicates that the recovery decline is largely associated with frequency-localization errors rather than incorrect source counting.

A likely explanation is the frequency-anchor procedure used by \SFG. Each source slot first selects a nearby spectral peak as a reference and then predicts the source frequency relative to that peak. When two slots select the same peak, one of the binaries may need a much larger frequency correction than the network was trained to predict reliably. For example, at a separation of $1000\,\mathrm{nHz}$, the required correction can exceed the characteristic scale $F=800\,\mathrm{nHz}$ used in the frequency parameterization. We also tested preventing slots from selecting an already used anchor, but applying this restriction without retraining reduced recovery further, from $0.73$ to $0.47$ at the largest separation.

These results identify a limitation of the current frequency-estimation procedure, where the network can correctly infer that two binaries are present while placing one at an inaccurate frequency. 

 \section{Bayes factors between source counts}
\label{app:bayes}

To assess whether the data favor one or two binaries, we compare the Bayesian evidence of the corresponding signal models. The evidence measures how well a model explains the observations, averaged over its allowed parameter values. A two-source model generally achieves a better fit because it has more free parameters, but the evidence also accounts for how much of that additional parameter space is supported by the data.

For a model containing $K$ binaries, the evidence is
\begin{equation}
Z_K=\int d\theta_{1:K}\,\pi(\theta_{1:K})\,\mathcal L(d\mid\theta_{1:K}),
\label{eq:evidence}
\end{equation}
where $\pi$ denotes the parameter prior and $\mathcal L$ the Gaussian-noise likelihood. We compare the models through the log Bayes factor $\ln B_{21}=\ln Z_2-\ln Z_1$, with positive values favoring two binaries.

Evaluating the evidence directly would require integrating over all eight parameters of each binary. We instead use a Laplace approximation~\cite{kassraftery}, which estimates the integral from the best-fitting parameters and the shape of the likelihood around them. For the high-SNR binaries considered here, the likelihood is strongly concentrated near its maxima, making this a computationally practical approximation.

We express the source parameters in coordinates where their simulation priors are uniform,
\begin{equation}
u=\left(
fT_{\rm obs},
\log_{10}\dot f,
\log_{10}\mathcal A,
\cos\iota,
\lambda,
\sin\beta,
\psi,
\phi_0
\right).
\end{equation}
The frequency prior covers the full analyzed band of $3\,983$ Fourier bins, allowing the one-source model to fit a binary anywhere in that band. The volume of the allowed parameter space for one source is denoted by $V$, with $\ln V=15.9$.

Near a best-fitting parameter vector $\hat u_K$, we approximate the log likelihood by
\begin{equation}
\ln\mathcal L(d\mid u_K)\simeq
\ln\mathcal L_{\max}
-\frac{1}{2}(u_K-\hat u_K)^\top
\Gamma_K(u_K-\hat u_K),
\end{equation}
where $\Gamma_K$ is the Fisher matrix of the $K$-source waveform model. In other words, the likelihood is approximated by a multidimensional Gaussian centered on the best fit, with the Fisher matrix describing how rapidly the likelihood decreases when the parameters are changed.

Integrating this Gaussian over an unrestricted parameter space gives the standard Laplace approximation for a single likelihood maximum\nhouba{~\cite{kassraftery,cornishlittenberg2007}},
\begin{equation}
Z_K\simeq
\frac{\mathcal L_{\max}}{V^K}
\frac{(2\pi)^{4K}}{\sqrt{\det\Gamma_K}}.
\end{equation}
The maximum likelihood measures how well the model fits the data, while the Gaussian volume measures the range of nearby parameter values that also provide good fits. Dividing by the prior volume $V^K$ accounts for the fraction of the allowed parameter space occupied by this region.

Some parameters may be only weakly constrained, so extending the Gaussian beyond their allowed prior ranges would overestimate the evidence. \nhouba{Writing the Gaussian volume in the Fisher eigenbasis, $(2\pi)^{4K}/\sqrt{\det\Gamma_K}=\prod_i\sqrt{2\pi/\Lambda_i}$, we limit the contribution of each eigendirection to the width available within the prior.} Including equivalent permutations of the source labels, our approximation becomes
\begin{equation}
\ln Z_K\simeq
\ln\mathcal L_{\max}
+\ln K!
-K\ln V
+\sum_i\ln\min\!\left(
\sqrt{\frac{2\pi}{\Lambda_i}},\,\ell_i
\right),
\label{eq:laplace}
\end{equation}
where $\Lambda_i$ are the eigenvalues of the Fisher matrix and $\ell_i=\sum_j|e_{ij}|L_j$ is the prior width projected onto the corresponding eigenvector $e_i$, with $L_j$ denoting the original prior widths. The minimum prevents weakly constrained directions from contributing more volume than the prior permits. The factor $K!$ accounts for equivalent permutations of the source labels under the labeled-parameter convention used for this approximation.

Finding a good fit for the one-source model is particularly important, since underestimating its maximum likelihood would artificially favor two binaries. We therefore optimize the one-source waveform from several initial guesses, including the parameters of either injected source, an amplitude-weighted combination, and frequencies between the two signals. Each optimization uses Gauss--Newton~\cite{nocedalwright} followed by Nelder--Mead~\cite{neldermead}, and we retain the solution with the highest likelihood.

The resulting evidences remain approximate. The Fisher matrix describes only the local likelihood shape, and limiting the Gaussian widths to the prior ranges does not fully account for correlations between these directions or multiple separated likelihood maxima. Varying the finite-difference steps used to construct the Fisher matrix changes the median $\ln B_{21}$ by less than $0.01$. An additional importance-sampling check gives a median difference of $5.4$ in $\ln B_{21}$, but its low effective sample sizes prevent a precise independent validation. We therefore use the Bayes factors to assess whether the observations favor two distinct binaries across the controlled experiments, without interpreting their numerical values as precise evidence measurements.

\section{Fisher references}
\label{app:fisher}

We estimate local parameter uncertainties using the Fisher information matrix,
\begin{equation}
\Gamma_{ab}=\left\langle\frac{\partial h}{\partial\vartheta_a},\frac{\partial h}{\partial\vartheta_b}\right\rangle,
\end{equation}
where $\vartheta$ denotes the waveform parameters and the noise-weighted inner product is defined in Eq.~\eqref{eq:inner}. In the linearized-signal approximation, the covariance is $\Gamma^{-1}$~\cite{cutlerflanagan,vallisnerifisher}. Derivatives are evaluated using central finite differences with steps of $10^{-3}$ times a characteristic parameter scale: $1/T_{\rm obs}$ for frequency, relative scales for amplitude and frequency derivative, and $1\,\mathrm{rad}$ for angles.

We calculate two uncertainty estimates. The \emph{conditional} Fisher covariance varies only the eight parameters of the binary of interest, assuming all neighboring binaries are known exactly. The \emph{joint-marginal} covariance instead allows all binaries to vary, using the corresponding source block of the inverse $8K\times8K$ Fisher matrix. The latter therefore includes the additional uncertainty arising from overlapping signals. Both estimates describe the local likelihood curvature around the injected parameters and do not capture multimodal posterior structure.

\section{Likelihood-refinement experiment details}
\label{app:handoff}

For the likelihood-based refinement, we use the same eight-dimensional likelihood for Source A, keeping the other six binaries fixed at their injected parameters.
All four initialization methods use the same affine-invariant ensemble MCMC sampler with the Goodman--Weare stretch move~\cite{goodmanweare}, at a single temperature and under a common budget of approximately $80\,000$ likelihood evaluations using the $A$, $E$, and $T$ channels. This corresponds to $48$ walkers advanced for $1\,701$ steps, or $81\,696$ evaluations including the initial walker positions. We repeat each initialization with $11$ independent random seeds. For the two reference initializations that use the injected parameters, the initial frequency spread is $10\,\mathrm{nHz}$ for the truth-centered walkers and $1.1\,\mathrm{nHz}$ for those drawn from the local Fisher covariance.

To measure how quickly the sampler finds a good fit, we first determine the highest log likelihood reached by any of the truth-initialized runs. We then record how many likelihood evaluations each run needs before its best walker reaches a log likelihood within one of this reference value. This provides a common measure of how quickly the different initializations locate a high-likelihood solution; it does not establish that the MCMC chains have converged or that they accurately sample the full posterior.

For Fig.~\ref{fig:handoff}(b), we measure how tightly the walkers concentrate around the recovered frequency as the likelihood search progresses. Some walkers may still be far from a good-fitting solution, so including all of them would make the frequency distribution appear much broader. We therefore select the $24$ walkers with the highest average log likelihood over their preceding $75$ steps and calculate the frequency interval between the $16$th and $84$th percentiles of their samples. The same procedure is applied to all four initialization methods. These widths describe the frequency spread of the better-fitting walkers, not necessarily the uncertainty of a converged posterior.
\bibliography{references}
\end{document}